\documentclass[10pt,preprint]{aastex6}
\usepackage{amsmath, amsthm}
\usepackage{booktabs}

\usepackage{comment}

\slugcomment{Submitted, \today}

\newcommand*{\Ledd}{L_{\rm Edd}}
\newcommand*{\calR}{{\cal R}}
\newcommand*{\calM}{{\cal M}}

\newcommand*{\Mdotedd}{\dot M_{\rm Edd}}
\newcommand*{\Mdotb}{\dot M_{\rm B}}
\newcommand*{\Mdotbe}{\dot M_{\rm e}}
\newcommand*{\Mdotbes}{\dot M_{\rm es}}

\newcommand*{\Mdott}{\dot M_{\rm t}}
\newcommand*{\Mdotacc}{\dot M_{\rm acc}}

\newcommand*{\lambdat}{\lambda_{\rm t}}

\newcommand*{\Lambdacr}{\Lambda_{\rm cr}}
\newcommand*{\lambdacr}{\lambda_{\rm cr}}
\newcommand*{\lambdaes}{\lambda_{\rm es}}

\newcommand*{\gmin}{g_{\rm min}}
\newcommand*{\fmin}{f_{\rm min}}
\newcommand*{\xmin}{x_{\rm min}}

\newcommand*{\Mbh}{M_{\rm BH}}
\newcommand*{\Mg}{M_{\rm g}}

\newcommand*{\cs}{c_{\rm s}}
\newcommand*{\rb}{r_{\rm B}}
\newcommand*{\rbe}{r_{\rm e}}
\newcommand*{\rg}{r_{\rm g}}

\newcommand*{\phit}{\phi_{\rm t}}
\newcommand*{\phig}{\phi_{\rm g}}

\newcommand*{\rhoinf}{\rho_{\infty}}
\newcommand*{\pinf}{p_{\infty}}
\newcommand*{\csinf}{c_{\infty}}
\newcommand*{\Tinf}{T_{\infty}}
\newcommand*{\rhotil}{\tilde\rho}
\newcommand*{\mpr}{m_{\rm p}}

\shortauthors{L. Ciotti and S. Pellegrini}
\shorttitle{Isothermal Bondi accretion in Jaffe and Hernquist galaxies}

\begin{document}

\title{Isothermal Bondi accretion in Jaffe and Hernquist 
  galaxies with a central black hole: fully analytical solutions}
 \author{Luca Ciotti$^{\star}$ and Silvia Pellegrini}
\affil{Department of Physics and Astronomy, University of Bologna,
  via Piero Gobetti 93/2, 40129 Bologna, Italy\\
$^{\star}$E-mail: luca.ciotti@unibo.it }


\begin{abstract}

  One of the most active fields of research of modern-day astrophysics
  is that of massive black hole formation and co-evolution with the
  host galaxy.  In these investigations, ranging from cosmological
  simulations, to semi-analytical modeling, to observational studies,
  the Bondi solution for accretion on a central point mass is widely
  adopted. In this work we generalize the classical Bondi accretion
  theory to take into account the effects of the gravitational
  potential of the host galaxy, and of radiation pressure in the
  optically thin limit. Then, we present the fully analytical
  solution, in terms of the Lambert-Euler $W$-function, for isothermal
  accretion in Jaffe and Hernquist galaxies with a central black hole.
  The flow structure is found to be sensitive to the shape of the mass
  profile of the host galaxy.  These results and the formulae that are
  provided, mostly important the one for the critical accretion
  parameter, allow for a direct evaluation of all flow properties, and
  are then useful for the above mentioned studies. As an application,
  we examine the departure from the true mass accretion rate of
  estimates obtained using the gas properties at various distances
  from the black hole, under the hypothesis of classical Bondi
  accretion.  An overestimate is obtained from regions close to the
  black hole, and an underestimate outside a few Bondi radii; the
  exact position of the transition between the two kinds of departure
  depends on the galaxy model.

\end{abstract}


\section{Introduction} 

Bondi (1952) presented the solution for spherically-symmetric, steady
accretion of a spatially infinite gas distribution on to an isolated
central point mass. This solution in the recent years became a
standard and intensively adopted tool for estimates of the
scale-radius and the mass flow rate in studies of accretion on massive
black holes (hereafter MBH) at the center of galaxies.  In fact, the
discovery of the ubiquitous presence of MBHs at the center of
spheroids on one side (Kormendy \& Richstone 1995), and the enormous
advance in instrumental capabilities, computer performances and numerical astrophysics on the
other, triggered a huge increase in the number of investigations
involving the accretion phenomenon at the center of galaxies.  These
studies range from observational works deriving the gas properties in
regions surrounding the MBHs (see, e.g., Baganoff et al. 2003;
Pellegrini 2005, 2010; Rafferty et al. 2006; Kormendy \& Ho 2013; Wong
et al. 2014; Russell et al. 2015), to theoretical studies on the
origin of the various types of AGN sources, and on the various
physical processes involved in accretion on to a MBH (e.g., Bu et
al. 2013, Yuan \& Narayan 2014, Cao 2016, Park et al. 2017), to
cosmological investigations of MBH formation, and co-evolution of MBHs
and host galaxies involving the ``feedback'' action (see, e.g.,
Sijacki et al. 2007; Hopkins et al. 2006, 2007; Di Matteo et al. 2008;
Park \& Ricotti 2011; Volonteri et al. 2015; Choi et al. 2016; DeGraf
et al. 2017). Unfortunately, in general these studies lack the resolution to follow gas
transport down to the parsec scale, and the Bondi model is used as
the starting point for estimates of the accretion radius (i.e., the
sonic radius), and of the mass accretion rate.  In particular, the
Bondi accretion rate gives the mass supply to the MBH by taking 
the density and temperature at some finite
distance from the center, implicitely assuming that these values
represent the true boundary conditions (i.e., at infinity) for the
Bondi problem.  Even the knowledge of the true boundary conditions, though,
would not be enough for a proper treatment of the real problem, 
because the MBH is not an isolated point-mass, but resides at the
bottom of the potential well of the host galaxy. Moreover, the
radiation emitted by the inflowing material may interact with it, 
accretion may become unsteady, and Bondi accretion during phases of AGN feedback 
cannot be applied (e.g., Ciotti \& Ostriker 2012).  During phases of moderate accretion
(in the ``maintainance'' mode), instead, when the problem can be considered almost
steady,  Bondi accretion could provide a reliable approximation
of the real situation.

In Korol et al. (2016, hereafter KCP16), we generalized the Bondi
problem to mass accretion at the center of galaxies, including also
the effect of electron scattering on the accreting gas. We then
calculated the deviation from the true values of estimates of the
Bondi radius and mass accretion rate due to adopting as boundary
values for the density and temperature those at a finite distance from
the MBH, and assuming the validity of the classical Bondi accretion
solution. In the case of the Hernquist galaxy model, we showed how to
obtain the analytical expression for the critical value of the
accretion parameter $\lambdacr$, for generic values of the polytropic
index $\gamma$. However, even for this exceptional case, the radial
profiles of the properties of the accreting gas remained to be
determined numerically.  Of course, in observational studies, or in
numerical simulations where subgrid MBH accretion is implemented,
analytical solutions for these radial profiles would be very
useful. Here we show that, remarkably, the whole accretion problem can
be solved in a {\it fully} analytical way for the {\it isothermal}
accretion in Jaffe (1993) and Hernquist galaxy models with central
MBHs. This is due to the fact that 1) for these two models it is
possible to obtain explicitely $\lambdacr$, and 2) in isothermal
accretion in generic spherically symmetric potentials the radial
profile of the Mach number can be explicitely written in terms of the
{\it Lambert-Euler} $W$-function. The possibility of using the
$W$-function to describe isothermal flows had been pointed out when
discussing the isothermal Parker solution for the solar wind [Cranmer
2004; see also Herbst (2015) for the case of accretion]. At the best
of our knowledge, the present work provides the first fully analytical
solution of the accretion problem on a MBH at the center of a galaxy.

This paper is organized as follows.  In Section 2 we recall the main
properties of the classical Bondi solution, and we present the fully
analytical isothermal solution for accretion onto an isolated MBH.  In
Section 3 we consider the generalized case of the Bondi solution in
presence of radiation feedback and of a galaxy potential hosting the
central MBH. Section 4 deals with the Jaffe and Hernquist
galaxy models, and presents the  fully analytical solution for them.
In Section 5 we examine the departure of the estimate of 
the  mass accretion rate from the true value, when the estimate is obtained 
using as boundary values for the density and
temperature those at points along the solution at finite distance from
the MBH. The main conclusions are summarized in Section 6,
and technical details are given in the Appendixes.

\section{The classical Bondi model} 
\label{sec:class}

We shortly recall here the main properties of the classical, polytropic Bondi
accretion model, even though the present investigation focusses on isothermal accretion.  The
gas is perfect, has a spatially infinite distribution, and is 
accreting on to an isolated central mass, in our case a 
MBH, of mass $\Mbh$. The gas density and pressure are linked by:
\begin{equation} 
p = {k_{\rm B} \rho T\over \mu\mpr} = \pinf\rhotil^{\gamma},\quad
\rhotil\equiv{\rho\over\rhoinf},
\end{equation}
where $1 \le \gamma \le 5/3$ is the polytropic index, $\mpr$ is the
proton mass, $\mu$ is the mean molecular weight, $k_{\rm B}$ is the
Boltzmann constant, and $\pinf$ and $\rhoinf$ are respectively the gas
pressure and density at infinity.  The polytropic gas sound speed is
\begin{equation}
\cs^2=\gamma {p\over\rho}.
\end{equation}
The isothermal case is recovered for $\gamma=1$.
The time-independent continuity equation is:
\begin{equation} 
4 \pi r^2 \rho(r) v(r)= \Mdotb,
\end{equation}
where $v(r)$ is the gas radial velocity, and $\Mdotb$ is the
time-independent accretion rate on the MBH.  The Bernoulli
equation, with the appropriate boundary conditions at infinity,
becomes:
\begin{equation}
{v(r)^2\over 2} + \Delta h(r) - {G\Mbh\over r} = 0,
\end{equation}
where, from eq. (1)
\begin{equation} 
\Delta h\equiv \int_{\pinf}^p {dp\over \rho} =\csinf^2\times
\begin{cases}
\displaystyle{
{\rhotil^{\gamma-1} -1\over\gamma-1} ,\quad \gamma>1,}
\\ \\
\displaystyle{\ln\rhotil ,\quad\quad\quad\gamma=1}, 
\end{cases}
\end{equation}
and $\csinf$ is the sound speed of the gas at infinity.

An important scalelength of the problem, the
so-called Bondi radius, is naturally defined as
\begin{equation} 
\rb \equiv {G\Mbh\over\csinf^2}.
\end{equation}
In the following  we will use $\rb$ as given by the definition above as the length scale, even when
considering more general models. After introducing the normalized quantities
\begin{equation}
x\equiv {r\over\rb},\quad
\tilde{\cs} \equiv {\cs\over\csinf}=\rhotil^{\gamma-1\over 2},\quad 
\calM={v\over\cs},
\end{equation}
where $\calM$ is the Mach number, 
eqs.  (3)-(4) become respectively
\begin{equation}
x^2 \calM \rhotil^{\gamma+1\over 2}={\Mdotb\over 4\pi\rb^2\rhoinf\csinf}\equiv\lambda,
\end{equation}
\begin{equation}
\begin{cases}
\displaystyle{
{\calM^2 {\tilde\cs}^2\over 2} + 
{\rhotil^{\gamma -1}\over \gamma -1} = {1\over x} +{1\over \gamma-1},\quad\gamma>1,}
\\ \\
\displaystyle{
{\calM^2 {\tilde\cs}^2\over 2} +\ln\rhotil = {1\over x},\quad\gamma =1.}
\end{cases}
\end{equation}
$\lambda$ is the dimensionless accretion parameter, that 
determines the accretion rate for assigned $\Mbh$ and boundary
conditions. From eqs. (7)-(8),  the radial
profile of all hydrodynamical properties can be expressed in terms of $\calM (x)$.  By elimination of $\rhotil$ in
eqs. (8)-(9), for $1<\gamma\leq 5/3$ the Bondi problem reduces to the
solution of the equation
\begin{equation} 
g(\calM)=\Lambda f(x), \qquad \Lambda \equiv
      \lambda^{2 (1-\gamma)\over \gamma +1},
\end{equation}
where
\begin{equation}
	g(\calM)\equiv \calM^{2 (1-\gamma)\over \gamma+1} 
                         \left( {\calM^2\over 2} + {1\over\gamma-1 }\right),
\end{equation}
and
\begin{equation}
	f(x)\equiv x^{4 (\gamma-1)\over\gamma +1}\left({1\over x}
         + {1\over \gamma-1}\right).
\end{equation}
As well known, $\Lambda$ cannot be chosen arbitrarily; in fact, both
$g(\calM)$ and $f(x)$ have a minimum:
\begin{equation} 
\gmin={\gamma+1\over 2(\gamma-1)},\qquad {\rm for} \quad \calM_{\rm min}=1, 
\end{equation}
and
\begin{equation} 
\fmin={\gamma+1\over 4(\gamma-1)}\left({4\over 5-3\gamma}\right)^{5-3\gamma\over\gamma+1},\quad {\rm for}\,\,
\xmin={5-3\gamma\over 4}.
\end{equation}
Therefore, to satisfy eq. (10) $\forall x > 0$
requires that $\gmin \leq \Lambda \fmin$, i.e., that $\Lambda \ge
\Lambdacr\equiv\gmin/\fmin$.  In order for the solution to exist it is then required that:
\begin{equation}
\lambda \leq \lambdacr = \left({\fmin\over\gmin}\right)^{\gamma+1\over
  2(\gamma-1)} = {1\over 4}\left({2\over 5-3\gamma}\right)^{5-3\gamma\over 2(\gamma-1)}.
\end{equation}
In particular $\lambdacr= e^{3/2}/4$ for $\gamma \to 1^+$, and
$\lambdacr=1/4$ for $\gamma \to {5/3}^-$. 

\begin{figure*}
\hskip -0.2truecm
\includegraphics[height=0.35\textwidth, width=0.35\textwidth]{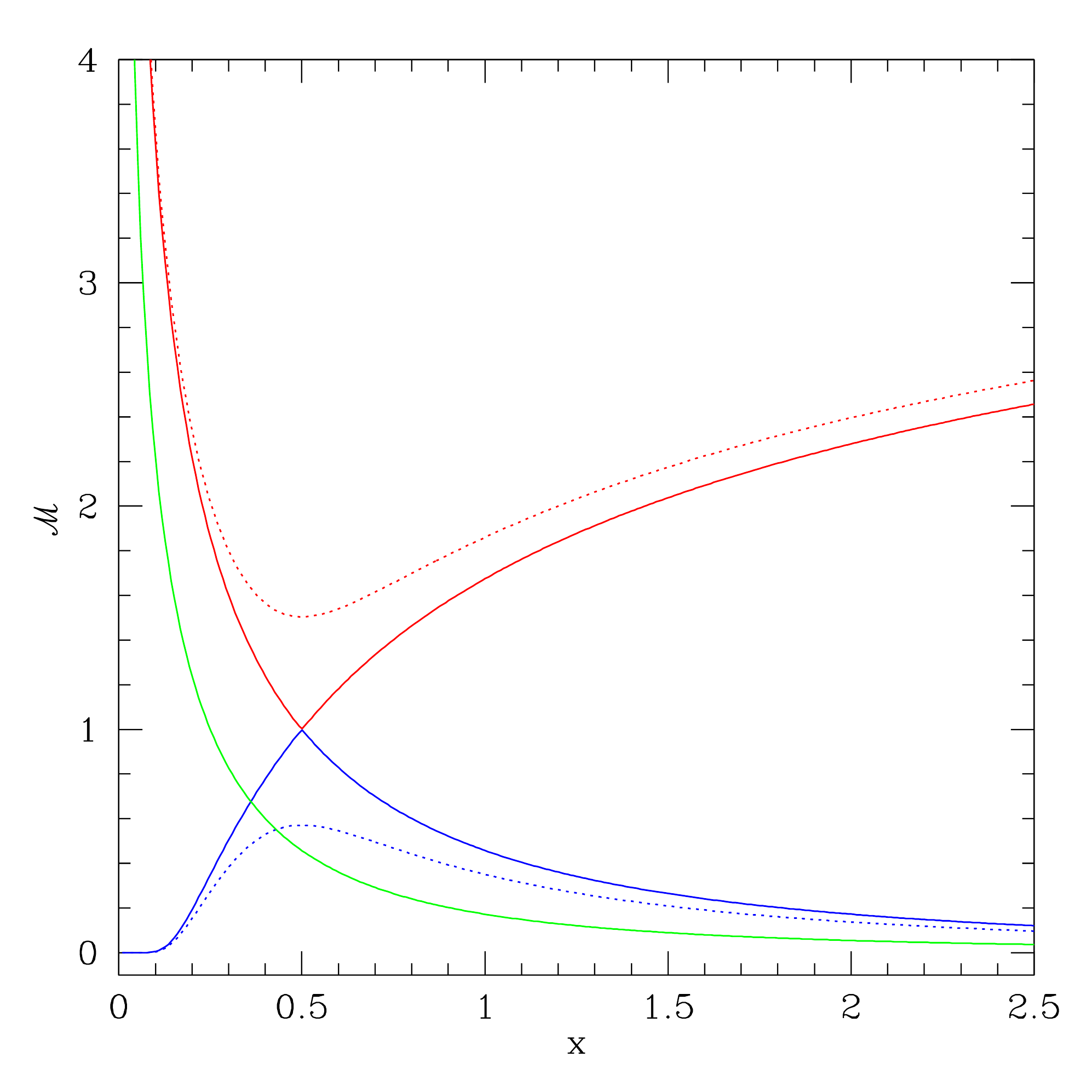}
\hskip -0.3truecm
\includegraphics[height=0.35\textwidth, width=0.35\textwidth]{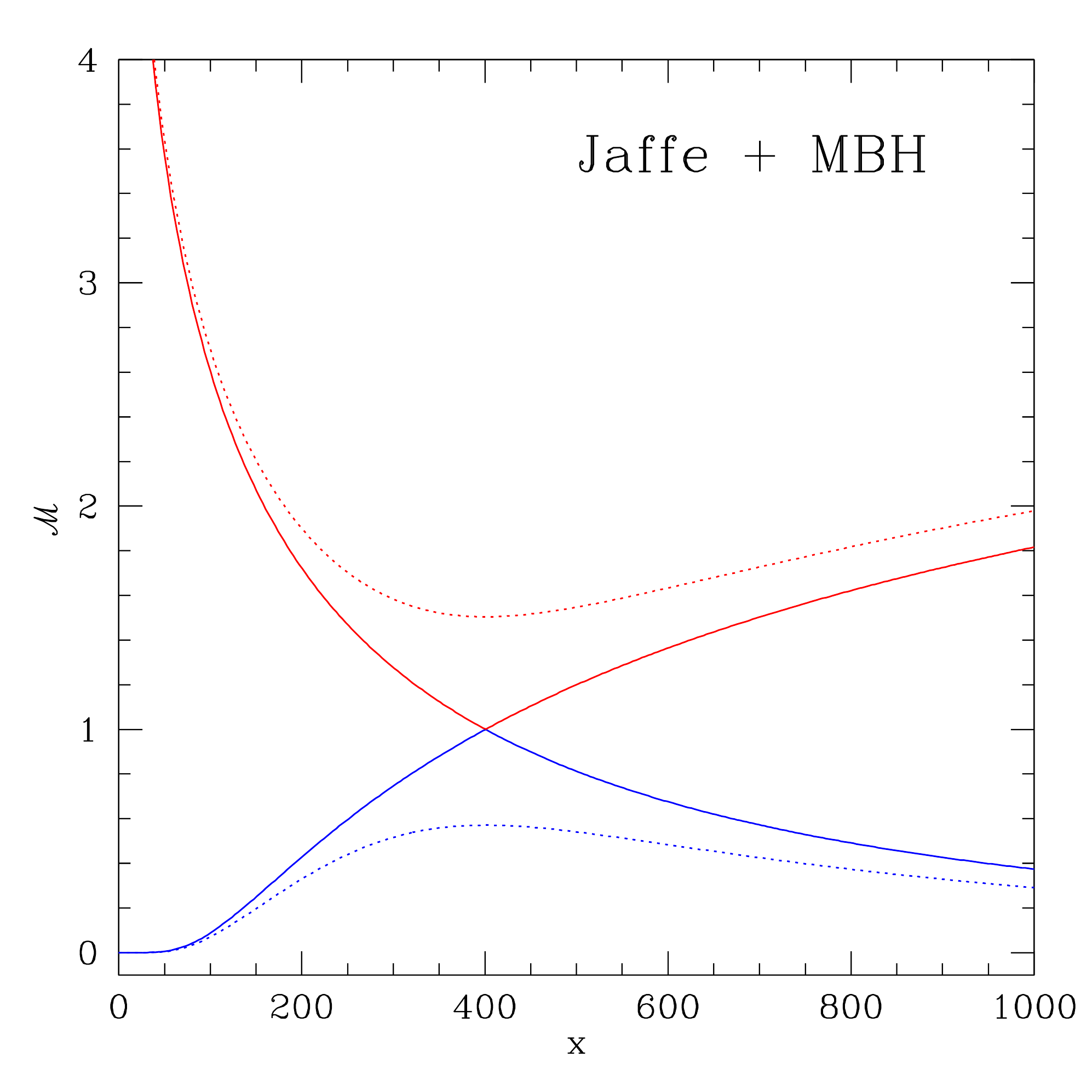}
\hskip -0.3truecm
\includegraphics[height=0.35\textwidth, width=0.35\textwidth]{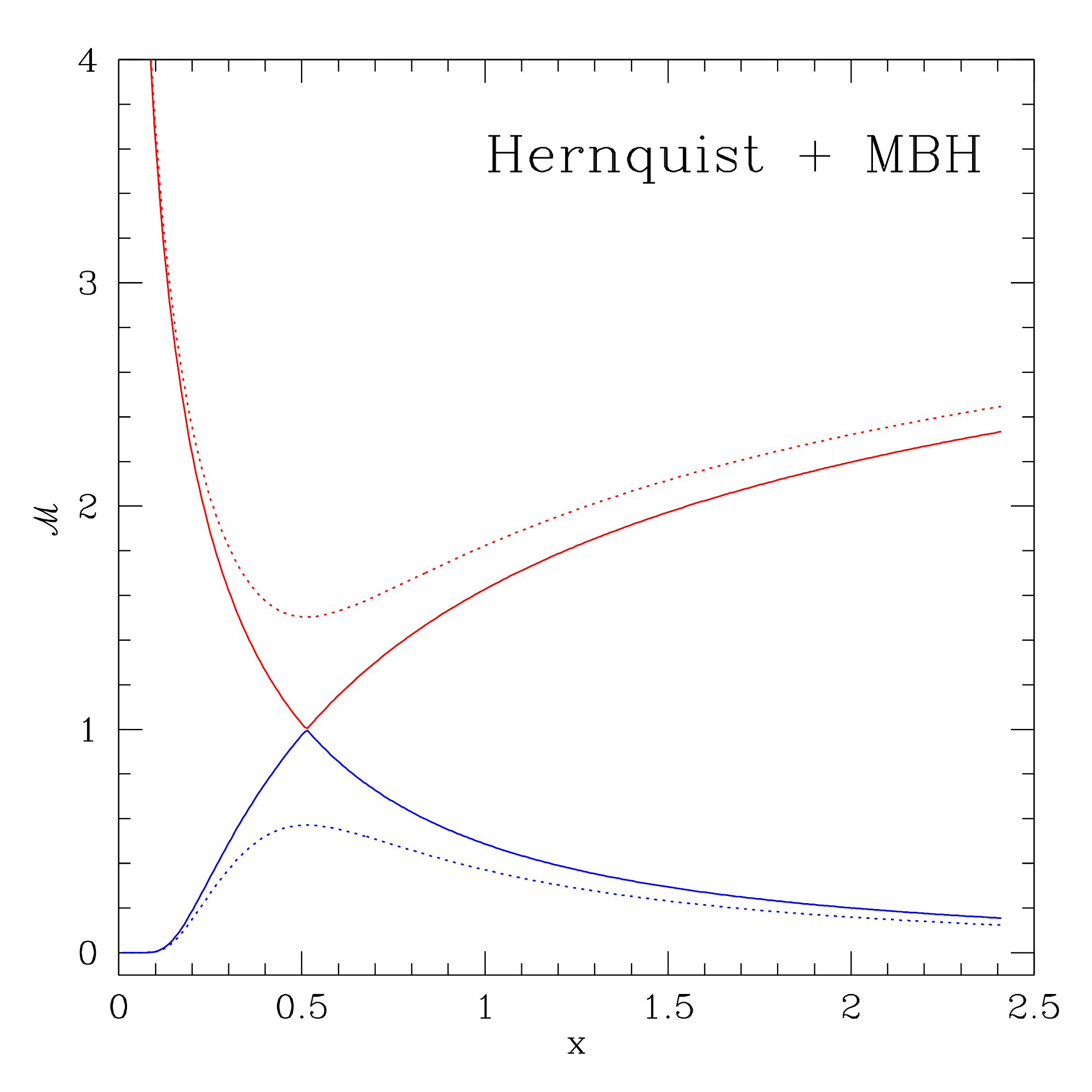}\\
\caption{Mach number profile $\calM(x)$ for isothermal Bondi accretion on a MBH. Blue lines show the subsonic regime,
red lines the supersonic one. The leftmost panel refers to classical Bondi
accretion (Sect. 2.1), and the green line refers to the inclusion of the effect of radiation pressure for $\chi=0.5$
(Sect. 3.1). The middle and right panels show respectively accretion at the center of a Jaffe and an 
Hernquist galaxy model, with $\calR=1000$, $\xi=100$ and $\chi=1$ (Sects. 4.1 and 4.2).
Solid lines show the two critical solutions, one in which $\calM$ continuously 
increases towards the center, and the other in which the flow 
starts supersonic and reaches the center with zero velocity. The dotted lines show the illustrative example 
of the two subcritical solutions with $\lambda=0.8\lambdacr$. 
Note how, for this particular choice of the $\calR$ and $\xi$ values,
the position of the sonic point moves considerably outward in presence of a Jaffe galaxy (Sect. 4.1); for the Hernquist one, instead, the sonic point is
similar to that of the classical Bondi accretion, that is it remains small (Sect. 4.2). }
\label{fig:machbh}
\end{figure*}

In the isothermal case the Bondi problem is given by:
\begin{equation}
\begin{cases} 
g(\calM)= f(x) -\Lambda,\qquad\Lambda \equiv \ln\lambda,
\\ \\
\displaystyle{g(\calM)\equiv {\calM^2\over 2}-\ln\calM ,}
\\ \\
\displaystyle{f(x)\equiv {1\over x}+2\ln x},
\end{cases}
\end{equation}
with 
\begin{equation}
\begin{cases}
\displaystyle{\gmin={1\over 2}, \qquad\qquad \quad {\rm for}\quad\quad \calM_{\rm min}=1,}
\\ \\
\displaystyle{\fmin= 2-2\ln 2, \qquad {\rm for} \quad\quad \xmin={1\over 2},}
\end{cases}
\end{equation}
so that solutions exist only for $\gmin \le \fmin -\Lambda$,
i.e., for $\Lambda\leq\Lambdacr\equiv\fmin-\gmin$. Equation (17) then 
requires
\begin{equation}
\lambda \leq \lambdacr = e^{\fmin-\gmin} ={{\rm e}^{3/2}\over 4},
\end{equation}
in accordance with the limit of eq. (15) for $\gamma\to 1^+$.  In the
following, the function $f(x)$ in eq. (16) is generalized by
considering the effect of radiation pressure due to electron
scattering, and the additional gravitational field of the host galaxy.

Summarizing, the solution of the Bondi problem requires to obtain the radial
profile $\calM(x)$, for given $\lambda \le \lambdacr$ (see, e.g.,
Bondi 1952; Frank, King \& Raine 1992). Unfortunately, eq. (10) does not have an explicit solution in
terms of known functions for generic values of $\gamma$, and must be
solved numerically.

\subsection{The isothermal solution}

Among the two critical solutions characterizing the Bondi problem, we
consider that pertinent to accretion, with an increasing Mach number approaching the center.
Appendix A shows how introducing the Lambert-Euler function $W$, and 
using eq. (A3) with $a=1/2$, $b=2$, $c=0$, $d=-1$, $X=\calM$ and
$Y=f(x)-\Lambda$, one can solve eq. (16) for $\calM(x)$, 
and then recover the full solution for  isothermal accretion. In particular, along the critical solution,
when $\lambda=\lambdacr$, $\calM ^2$ is given by:
\begin{equation}
\calM^2= -
\begin{cases}
\displaystyle{
W\left(0,-{{\rm e}^{3-2/x}\over 16x^4}\right),\quad x\geq 
\xmin},
\\ \\
\displaystyle{
W\left(-1,-{{\rm e}^{3-2/x}\over 16x^4}\right),\quad 0< x\leq 
\xmin,}
\end{cases}
\end{equation}
where $\xmin =1/2$ [eq. (17)].

The properties of the critical solution (19) can be visualized with the help of Fig. 5 in Appendix A.
Figure 5 shows that only for a negative argument $W$ assumes negative values, and so
$\calM^2$ can be a positive quantity.  As $x$ decreases from infinity to
the sonic point $\xmin =1/2$, the argument $z$ of the $W(0,z)$
function decreases from $0$ to $-1/e$, and the function $W(0,z)$
decreases from 0 to $-1$, i.e. from point $A$ to point $B$ (solid line in
Fig. 5). Then, when $x$ decreases further approaching the origin, the
argument of the branch $W(-1,z)$ increases again from $-1/e$ to $0$,
and $W$ decreases from $-1$ to $-\infty$, moving from
point $B$ to the asymptotic point $C$ (dashed line in Fig. 5). Note that the $W(-1,z)$ function describes
supersonic accretion, while $W(0,z)$ subsonic accretion, so that the critical
solution is obtained by connecting together the two
branches. This is illustrated by the red and blue solid lines, respectively, in Fig. 1. Note also that, in case of isothermal accretion with
$\lambda <\lambdacr$, the continuous solutions are limited to a subset of the
regions $A$-$B$ (the subsonic accretion branch, blue dotted lines), or to a subset of the region $B$-$C$ (the
supersonic accretion branch, red dotted lines).

Having obtained $\calM(x)$, from eq. (8) with $\gamma=1$ one has:
\begin{equation}
\rhotil(x)={\lambdacr\over x^2\calM (x)},
\end{equation}
while the modulus of the accretion velocity is $v(r)=\csinf\calM(x)$.

\section{Generalized Bondi accretion for an isothermal flow}

\subsection{Adding the effect of electron scattering}
\label{sec:escatt}

The Bondi solution describes a purely hydrodynamical flow, where heat
exchanges are implicitly taken into account by the polytropic index\footnote{
$\gamma$ is not necessarily the adiabatic index $\gamma_{ad}$, so that in
the Bondi solution the entropy of the gas can change along the radial
streamlines. For a polytropic transformation,
of specific heat at constant volume ${\cal C}_V$, the molar specific heat is ${\cal C}={\cal C}_V
 (\gamma_{ad}-\gamma)/(1-\gamma)$ (e.g., Chandrasekhar
 1939). Therefore ${\cal C}<0$ when $1<\gamma<\gamma_{ad}$, and the fluid
 element loses energy as it moves inward and heats.} (see also KCP16,
Sect. 3).
In real cases the flow can be affected by the  emission of radiation
near the MBH; in terms of the mass accretion rate $\Mdotacc$, this is expressed by 
\begin{equation} 
	L= \varepsilon\Mdotacc c^2,
\end{equation}
where $\varepsilon$ is the radiative efficiency. For example, in the
classical Bondi accretion one would adopt $\Mdotacc=\Mdotb$. The efficiency
$\varepsilon$ can depend on $\Mdotacc$, as in the advection dominated accretion at 
low $\Mdotacc$  (e.g., Yuan \& Narayan 2014), when $\varepsilon$ is very low.  At high accretion rates, instead,
the efficiency is $\varepsilon_0\simeq 0.1$, and the effects of the
emitted radiation can be sufficiently strong to stop accretion; in these
circumstances, the steady  Bondi solution  cannot be applied, even approximately
(e.g., Ciotti \& Ostriker 2012 for a review).

KCP16 extended the classical Bondi accretion solution by including
 the radiation pressure effect due to 
electron scattering in the optically thin regime (see also Fukue
2001; Lusso \& Ciotti 2011).  Under the assumption of spherical
symmetry, the total (gravity plus radiation) force on a
gas element is:
\begin{equation}
	F(r)= -{G \Mbh \rho(r)\chi\over r^2},\quad 
        \chi\equiv 1-l,\quad 
         l \equiv {L\over\Ledd},
\end{equation}
where $\Ledd=4\pi c G \Mbh\mpr /\sigma_{\rm T}$ is the Eddington
luminosity, and $\sigma_{\rm T}=6.65 \times 10^{-25}$cm$^2$ is the
Thomson cross section. When $L=\Ledd$, and then $\chi =0$, radiation pressure cancels
exactly the MBH gravitational field at all radii; when $\chi=1$, the radiation pressure has no effect on the accretion flow.
We define $\Mdotbes$ as the accretion rate under
the above hypotheses; then, in terms of the Eddington mass accretion rate $\Mdotedd $, one has:
\begin{equation}
\Mdotedd\equiv{\Ledd\over \varepsilon_0c^2},\quad l = {\Mdotbes\over\Mdotedd} {\varepsilon\over\varepsilon_0}. 
\end{equation}
As demonstrated by KCP16, the value of $\Ledd$ to be inserted in eqs. (22) and (23)
remains that corresponding to the gravity produced only by the MBH, even when the gravitational potential of 
the host galaxy is also present.

The radiation feedback can be implemented as a correction that
reduces the effective gravitational force of the MBH by the factor
$\chi$. In particular, for $\gamma=1$ the function $f$ in eq. (16) becomes:
\begin{equation} 
	f(x)={\chi\over x} + 2\ln x,
\end{equation}
and the 
position of the minimum of the new $f$ moves inward with respect to 
the classical case:
\begin{equation} 
\xmin ={\chi\over 2}. 
\end{equation}

From the value of $\xmin$ and $\fmin =f(\xmin)$, one derives the critical value of the accretion
parameter, that we now call $\lambdaes$:
\begin{equation}
	\lambdaes = \chi^2\lambdacr ,
\end{equation}
where $\lambdacr$ is the critical parameter in the corresponding
classical case. In turn, the new accretion rate, $\Mdotbes$, is reduced with respect to $\Mdotb$, for given $\Mbh$ and
boundary conditions at infinity:
\begin{equation} 
\Mdotbes = 4 \pi \rb^2 \lambdaes\rhoinf\csinf = {\lambdaes\over\lambdacr}\Mdotb=\chi^2 \Mdotb. 
\end{equation}
The equation above is actually an implicit  equation for $\Mdotbes$, because 
$\chi$ depends on $\Mdotbes$ itself through eqs. (22)-(23). It can be shown that
$\Mdotbes$ self-regulates, and cannot exceed $\Mdotedd$, regardless of the size of 
$\Mdotb$ (see KCP16 for a more detailed discussion).

We then use the same procedure of Sect. 2.1 to derive 
the Mach profile of the critical isothermal solution, for generic values of $\chi$; $\calM (x)$ is now given by:
\begin{equation}
\calM^2= -
\begin{cases}
\displaystyle{
W\left(0,-{\chi^4{\rm e}^{3-2\chi/x}\over 16x^4}\right),\quad x\geq 
\xmin},
\\ \\
\displaystyle{
W\left(-1,-{\chi^4{\rm e}^{3-2\chi/x}\over 16x^4}\right),\quad 0< x\leq 
\xmin ,}
\end{cases}
\end{equation}
where now $\xmin = \chi/2$.  In Fig. 1 (left panel), the
green line compares the $\calM$ profile of the critical solution with that of the classical Bondi
problem, for the illustrative case $\chi=0.5$; as dictated by
eq. (25), the sonic radius moves inward, with $\xmin=1/4$. 
The density profile, setting $\gamma=1$ in eq. (8), becomes:
\begin{equation}
\rhotil(x)={\lambdaes\over x^2\calM (x)}. 
\end{equation}

\subsection{Adding the potential of the galaxy}
\label{sec:Bondi+gal}

We now move to the most general problem of Bondi accretion with electron scattering in the 
potential of the galaxy hosting the MBH at its center.  We write the gravitational potential of a
spherical galaxy of finite total mass $\Mg$ as
\begin{equation}
\phig = - {G\Mg\over\rg}\,\psi\left({r\over\rg}\right),
\end{equation}
where $\rg$ and $\psi$ are respectively a
characteristic scale-length, and the dimensionless galaxy potential.
We assume that $\psi =0$ for $r\to
\infty$, so that the Bernoulli constant in eq. (4) can still be fixed
at infinity. We further introduce the parameters
\begin{equation} 
\calR \equiv {\Mg\over\Mbh},\quad 
\xi \equiv {\rg\over\rb},
\end{equation}
with which the effective total potential becomes:
\begin{equation} 
\phit = - {G\Mbh\over\rb} \left[{\chi\over x} + {\calR\over\xi}\, \psi\left({x\over\xi} \right)\right].
\end{equation} 
The function $f(x)$ in eq. (16) is now written as:
\begin{equation}
 f={\chi\over x}+
{\calR\over\xi}\psi\left ({x\over\xi}\right) +2\ln x.
\end{equation}
Therefore the values of $\xmin$, $\fmin$, and of the critical
accretion parameter (that now we call $\lambdat$) depend on 
$(\chi,\calR,\xi)$, while the function $g(\calM)$ is unaffected by the
addition of the galaxy potential. Note how, for $\calR\to 0$ (or
$\xi\to\infty$), the galaxy contribution vanishes,
$\lambdat=\lambdaes$, and one recovers the solution of Sect. 3.1.
In the limiting case of $\chi=0$, the problem reduces to Bondi accretion
in the potential of the galaxy only, without electron
scattering and MBH.

The presence of a galaxy changes the accretion rate, that we now call
$\Mdott$; eq. (8) still holds, with:
\begin{equation} 
	\Mdott = 4 \pi \rb^2 \lambdat \rhoinf
      \csinf={\lambdat\over\lambdacr}\Mdotb,
\end{equation}
where again $\Mdotb$ is the classical Bondi accretion rate for the same chosen
boundary conditions $\rhoinf$ and $\csinf$.

The radial trend of the Mach number for the critical solution can be
derived again along the lines described in Sect. 2.1; we now have:
\begin{equation}
\calM^2= -
\begin{cases}
\displaystyle{
W\left(0, - {\rm e}^{-2f+2\Lambda_{\rm t}}\right), \quad x\geq \xmin 
}
\\ \\
\displaystyle{
W\left(-1, - {\rm e}^{-2f+2\Lambda_{\rm t}}\right), \quad 0< x\leq \xmin 
}
\end{cases}
\end{equation}
where $\Lambda_{\rm t} = \ln\lambdat = \fmin - \gmin$ as explained in Sect. 2.  Finally,  we have 
\begin{equation}
\rhotil(x)={\lambdat\over x^2\calM (x)}. 
\end{equation}

The full solution is then known, provided $\xmin$ is known.
For a generic galaxy model, $\xmin$, $\fmin$, and $\lambdat$ can be
determined only numerically. Moreover, the galaxy potential can
produce more than one minimum for $f(x)$ (KCP16).  As we will see in
the next Section, two of the most common galaxy models remarkably
allow for an analytical expression for $\xmin$ and $\Lambda_{\rm t}$, in the
isothermal case; therefore, {\it together with the expression above for} $\calM $, we have a 
{\it fully analytical solution of isothermal accretion on MBHs at the center of such galaxies}.

Note that along the critical solution the argument of the exponential in eq. (35) is $-2(f -\fmin) -1$; 
thus, as $x$ decreases from infinity to $\xmin$, the argument $z$ decreases from 0 to $-1/e$, and
the $W$-function moves from $A$ to $B=(-1/e,-1)$ in Fig. 5 of Appendix A, corresponding to 
the sonic point. Decreasing $x$ further, the solution switches on the $W(-1,z)$ branch, and
the Mach number diverges at the center. For $\Lambda <
\Lambda_{\rm t}$, the subsonic and supersonic cases are again described respectively
by the $W(0,z)$ and $W(-1,z)$ branches.

\section{Two fully analytical cases: Jaffe and Hernquist galaxy models
with a central MBH} 
\label{sec:Hernquistgal}
In this Section we present the final goal of our investigation, the 
analytical solution for all quantities describing accretion, in the isothermal case, for two particular 
galaxy models, the Hernquist (1990) and the Jaffe (1983) models with a central MBH.
These density profiles describe well the mass distribution of
early-type galaxies; they belong to the widely used family of the
so-called $\gamma$-models (Dehnen 1993, Tremaine et al. 1994), that
all have similar $\sim r^{-4}$ density profiles in their external regions (outside $\rg$).
For example, the Hernquist model with a central MBH has been recently adopted for a
numerical investigation of bulge-driven, Bondi fueling of seed black
holes (Park et al. 2016).
As already noticed, $\lambdat$ is known once the {\it absolute}
minimum $\fmin(\chi,\calR, \xi)$ is known; this, in turn, requires the
determination of $\xmin (\chi,\calR,\xi)$. 
Quite remarkably, for the Hernquist model $\xmin$ can be derived analytically in the general
polytropic case, solving a cubic equation (KCP16). A peculiar property of such equation is the possibility to present {\it
  two} minima for $x>0$, depending on the galaxy parameters; KCP16 provided the formulae needed to determine
the critical points of $f$,  for any choice of $\gamma$ and of the galaxy parameters, but the final
expressions for $\xmin$ and $\fmin$ were not given. Here we give these
analytical expressions for $\xmin$ and $\fmin$, in case of isothermal
accretion. In addition, we show that $\lambdat$
can be evaluated explicitely also for isothermal accretion in the Jaffe model; 
to our knowledge, these are the only two known cases of fully solvable accretion problems.  As we
will see, the Jaffe case turns out to be simpler than the Hernquist
one; this is not unexpected, because the Jaffe gravitational
potential is logarithmic, like the term appearing
in the expression for $f$ in eq. (33). In practice, $f$ has only one minimum, for all values
of the galaxy parameters. For this reason we begin the discussion with
the Jaffe model.

\subsection{The Jaffe model}

The gravitational potential of the Jaffe (1983) density distribution is given by 
\begin{equation} 
	\phig={G\Mg\over\rg}\ln\left({r\over r+\rg}\right),
\end{equation}
where the scale-length $\rg$ is related to the galaxy effective radius
$R_{\rm e}$ as $\rg \simeq R_{\rm e}/0.7447$. 
From eq. (37), $f$ in eq. (33) becomes: 
\begin{equation} 
	f = {\chi\over x}  - {\calR\over\xi}\ln\left({x\over x+\xi}
       \right) +  2\ln x.
\end{equation}

In three cases the determination of $\xmin$ and $\fmin$ is trivial. First, 
when $\xi \to \infty$ (or
$\calR \to 0$), the galaxy contribution vanishes, and the results in
Sect. 3.1 are recovered. Second, although
$\phig$ is not continuous at the origin, by fixing $r>0$ and
considering the limit $\rg\to 0$,  the
resulting potential is that of a MBH of mass $\Mg$; thus, this
limiting case reduces to the problem of accretion on a MBH of total
(effective) mass $(\chi+\calR)\Mbh$, and can be treated as in Sect. 3.1.
In particular, the position of the only minimum of $f$
(i.e., of the sonic radius), and the critical value $\lambdat$, are
given by:
\begin{equation} 
\xmin = {\chi+\calR\over 2},\quad \lambdat =
(\chi+\calR)^2 \lambdacr,
\end{equation}
where $\lambdacr$ is the critical value for the corresponding
classical Bondi problem (Sect. 2).  Third, for $\chi=0$ the discussion in
Appendix B shows that isothermal accretion in Jaffe potential is
possible only for $\calR\geq 2\xi$; this condition is equivalent to the
requirement that $G\Mg\geq2\rg c_{\infty}^2$.
We derive for $\xmin$ and $\lambdat$ the expressions:
\begin{equation}
\xmin ={\calR -2\xi\over 2}, \quad \lambdat ={\calR^2 \over {4\sqrt{e}}}\left( 1-{2\xi\over \calR} \right) ^{2-\calR/\xi}.
\end{equation}
This is equivalent to have $r_{\rm min}/\rg = G\Mg/(2\csinf^2\rg)-1$; for $\calR=2\xi$, one has that 
$\xmin=0$ and $\lambdat=\xi^2/{\sqrt{e}}$.

The general expression for $\xmin$, for assigned $\chi$, $\calR$, and
$\xi$, is given in eq. (B2); from this expression, $\fmin$ can be
easily evaluated. In Appendix B we also show that there is only one
minimum at $x\geq 0$. Figure 2 shows the trend of $\xmin$ and
$\lambdat$ with $\xi$, for representative values of $\calR$, and $\chi
=1$.  For the ease of inspection of the figure, we recall that the
choice of $\calR=1000$ corresponds to the standard assumption about
the ratio between the stellar mass and the MBH mass in spheroids
(e.g. Magorrian et al. 1998); a value of $\xi\approx 100$ is
representative of the case of hot gas accretion at the center of
elliptical galaxies, when $r_{\rm g}$ is a few kpc, and $r_{\rm B}$ is
of the order of a few tens of pc (e.g. Pellegrini 2010; see also
Sect. 6).  Figure 2 shows that $\xmin$ for accretion in the Jaffe
potential is much larger than in the classic Bondi solution ($\xmin
=1/2$); for reasonable $\xi \approx 100$ and $\calR \approx 10^3$ (see
above), $\xmin$ reaches values of a few hundreds. This significant
increase is explained by the compactness of the mass distribution that
is typical of the Jaffe model. The critical $\lambdat$ is also much
larger than in the classical Bondi accretion (Sect. 2).  In the
limiting case of $\xi\to 0$, the sonic radius $\xmin$ becomes very
large in a way predicted by eq. (39); this same equation also explains
the trend for $\lambdat\sim \calR^2$ shown at low $\xi$ in the right
panel of Fig. 2.  At large $\xi$, instead, the effect of the
gravitational field of the galaxy diminishes, and correspondingly
$\xmin$ decreases, towards the limiting case of classical Bondi
accretion; the same is true for $\lambdat$, that tends to $\lambdacr$.
Finally, cases with $\chi <1$ are not shown in Fig. 2, since their
differences with the plotted curves are very small, and would be
appreciated only for very large values of $\xi$ (or small values of
$\calR$), i.e. if the gravitational field at the sonic radius is
dominated by the presence of the MBH (see eq. 33).

The critical isothermal solution is given by
\begin{equation}
\calM^2= -
\begin{cases}
\displaystyle{
W\left[0, - {{\rm e}^{2\fmin -{2\chi\over x}}\over {\rm e}\,x^4}
\left({x\over x+\xi}\right)^{{2\calR\over\xi}}\right], \quad x\geq \xmin ,
}
\\ \\
\displaystyle{
W\left[-1, - {{\rm e}^{2\fmin -{2\chi\over x}}\over {\rm e}\,x^4}
\left({x\over x+\xi}\right)^{{2\calR\over\xi}}\right], 
\quad\quad 0< x\leq \xmin .
}
\end{cases}
\end{equation}
An illustrative solution for the above $\calM (x)$ 
is given in Fig. 1 (middle panel), for the typical values of $\calR=1000$ and $\xi=100$, and for $\chi=1$.
The choice of $\chi=1$ means no effect from the radiation pressure, thus Fig. 1 (as Fig. 2 just discussed)
shows purely gravitational effects due to the galactic + MBH potentials; however, as is the case for
$\xmin$ and $\lambdat$ in Fig. 2,  the differences in the trend of $\calM$
for $\chi < 1$ from those shown in Fig. 1 would be noticed only for very large $\xi$, or very small $\calR$.

\begin{figure*}
\hskip 0.3truecm
\includegraphics[height=0.5\textwidth, width=0.5\textwidth]{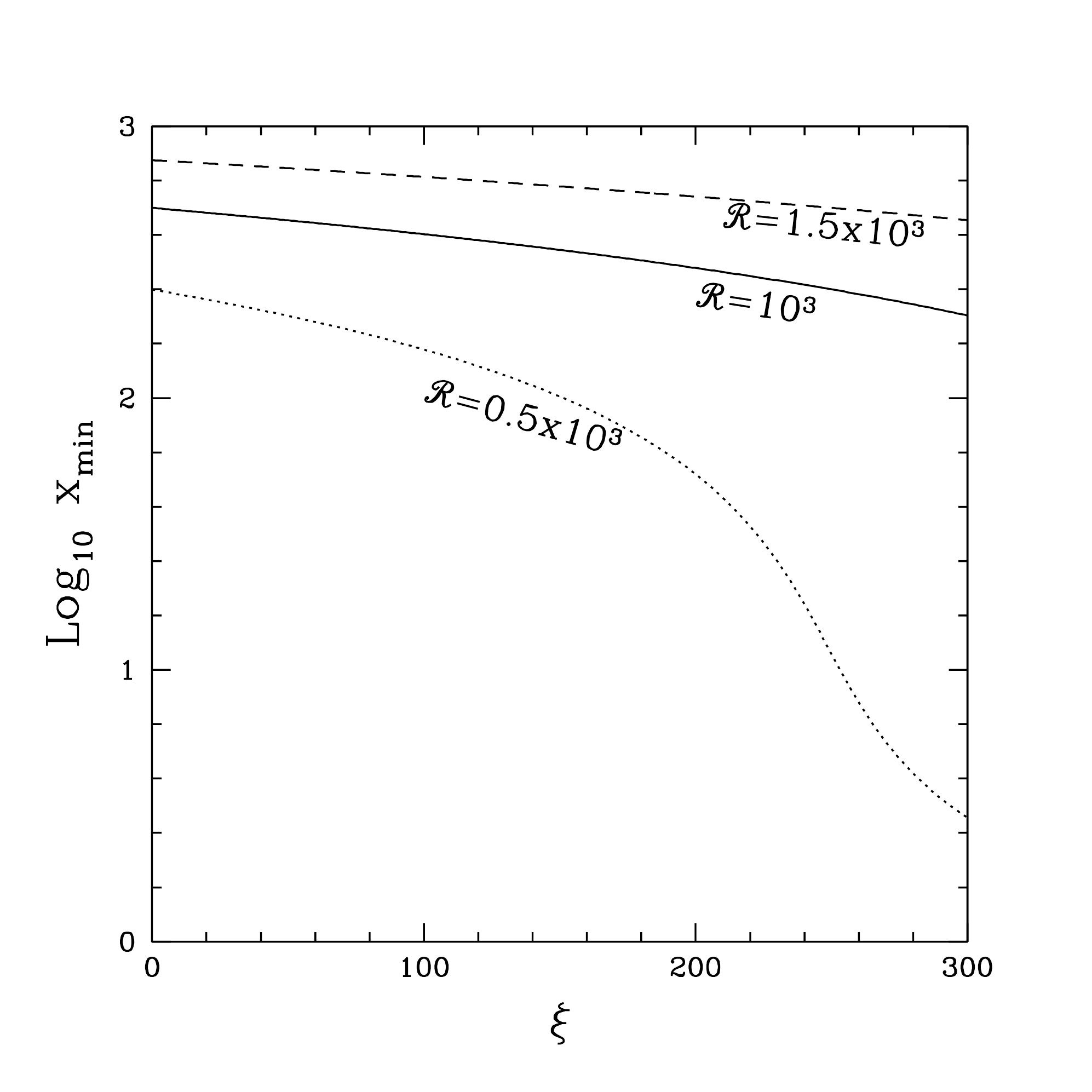}
\includegraphics[height=0.5\textwidth, width=0.5\textwidth]{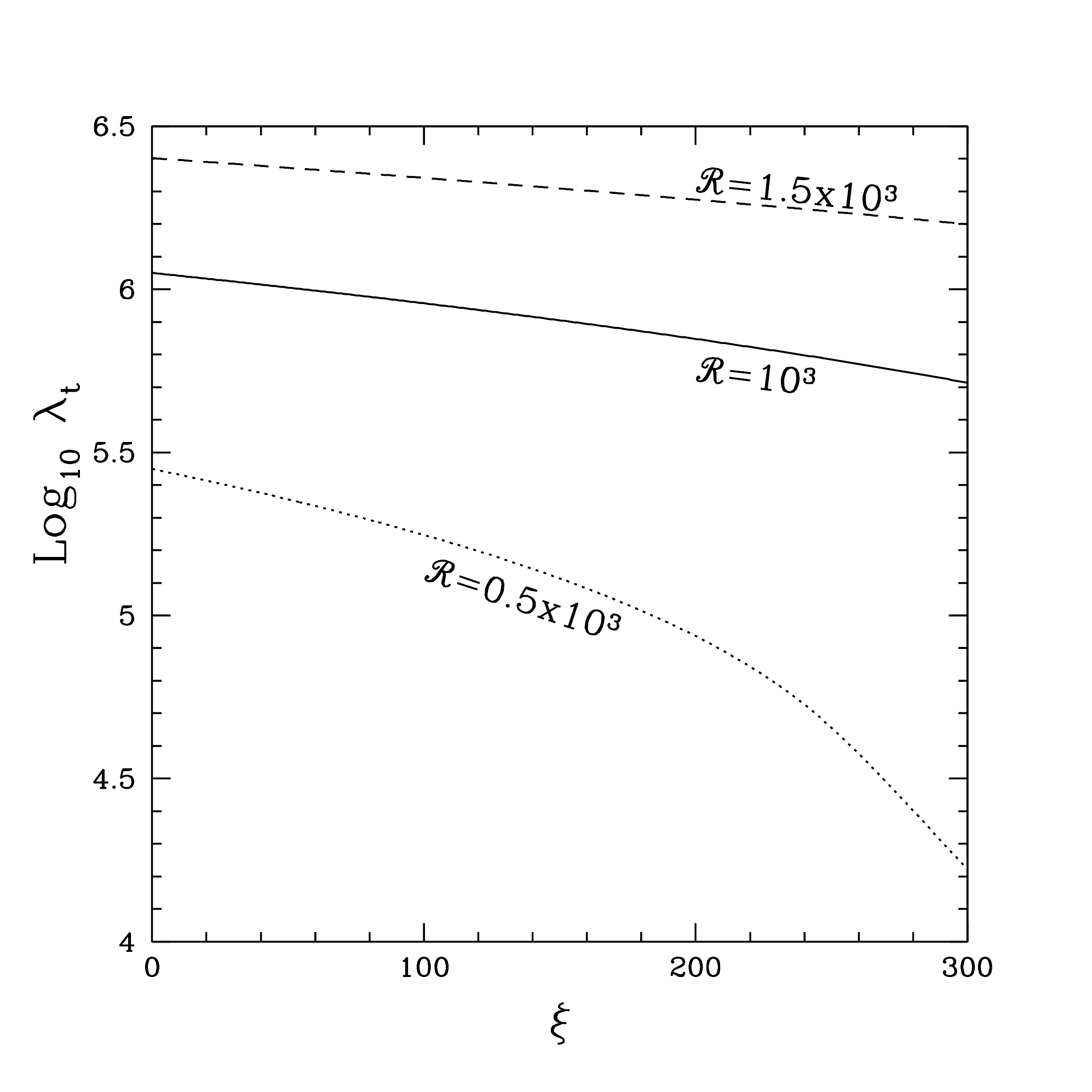}\\
\caption{Left: Position of the sonic point for isothermal accretion in
 a Jaffe galaxy model, as a function of $\xi$. The curves have been determined using eq. (B2),
and refer to a galaxy mass-to-MBH mass ratio of $\calR=500$
 (dotted), $1000$ (solid), $1500$ (dashed); $\chi=1$ is assumed.  Right: the corresponding values of
 $\lambdat$, determined following eq. (18), with $g_{\rm min}=1/2$ and with $f_{\rm min}$ calculated using eq. (38) for $x=x_{\rm min}$.  }
 \label{fig:jaffexmin}
\end{figure*}

\subsection{The Hernquist model}

The gravitational potential of the Hernquist model is:
\begin{equation} 
	\phig=-{G\Mg\over r+\rg},
\end{equation}
where $\rg \simeq R_{\rm e}/1.82$; thus, from eq. (33) one has 
\begin{equation} 
	f = {\chi\over x} + {\calR\over x+\xi} +  2\ln x .
\end{equation} 

As for the Jaffe model, from the expression above 
one sees that in three cases the determination of $\xmin$ and
$\fmin$ is trivial.  First, when $\xi \to \infty$ (or $\calR \to 0$),
the galaxy contribution to $f$ vanishes, and the results of Sect. 3.1 are
recovered. Second, for $r>0$ and $\rg\to 0$, $\phig$ reduces to the
potential of a MBH of mass $\Mg$, and the solution of Sect. 3.1 
applies, for a MBH of total (effective) mass $(\chi+\calR)\Mbh$, and
eq. (39) gives the position of $\xmin$. Third, for $\xi >0$ and $\chi =0$, at variance
with the case of the Jaffe model, 
no accretion is possible, because the minimum of $f$ is attained for $x\to 0^+$,
with $\fmin\to -\infty$, and so formally $\lambdat=0$ is needed. Therefore, from
now on we assume $\xi >0$ and $\chi >0$.

The procedure to determine  $\xmin$ analytically, for assigned $\chi$, $\calR$, and
$\xi$, is given in Appendix B. More than one sonic point can be present; the search for the absolute minimum
$\xmin$ can then be difficult. In turn,  the profile $\calM(x)$ of the critical solution can be non monotonic. 
Once $\xmin$ is determined, one can derive $f_{\rm min}$, and then evaluate $\lambda_{\rm t}$.  
The critical isothermal solution is then:
\begin{equation}
\calM^2= -
\begin{cases}
\displaystyle{
W\left(0, - {{\rm e}^{2\fmin -{2\calR\over x+\xi} -{2\chi\over
      x}}\over {\rm e}\,x^4}\right), \quad x\geq \xmin 
},
\\ \\
\displaystyle{
W\left(-1, - {{\rm e}^{2\fmin -{2\calR\over x+\xi} -{2\chi\over
      x}}\over {\rm e}\,x^4}\right), 
\quad 0< x\leq \xmin.
}
\end{cases}
\end{equation}

Representative trends for $\xmin$ and $\lambdat$ are plotted in Fig. 3, as a function of $\xi$,
for three values of $\calR$, and  for $\chi =1$. The parameters are the same chosen for the Jaffe model in Fig. 2,
but the trends are very different; the difference is due to the fact that 
there can be one or  three critical points for $f$ (two minima and one maximum), as discussed in Appendix B.
The curves in Fig. 3 refer to the absolute minimum. This different behavior is quite surprising, since both the Jaffe and Hernquist 
models belong to the class of $\gamma-$models, and are quite similar (for example, in projection they are both good 
approximations of the de Vaucouleurs surface brightness profile, over  a large radial range). The only major difference 
is that the model density in the central regions goes as $r^{-2}$ in the Jaffe acse, and as $r^{-1}$ in the Hernquist one.

The left panel shows how $\xmin$ can vary largely.
The easiest way to understand this trend is to refer to Fig. 6 in Appendix B, at a fixed $\calR$, and moving from left
to right. First we consider the black lines in Fig. 3, left panel, that populate small and large $\xi$ values; in these cases,
there is a {\it single} minimum for $f$. In fact, for given $\calR$ and sufficiently small or large $\xi$, there is 
a single minimum, according to Fig. 6. In particular, for very small $\xi$, the only minimum is placed at very large radii,
in accordance with eq. (39); this explains the large values of $\xmin$ in Fig. 3. Increasing $\xi$, the depth of the galactic 
potential well becomes shallower, and $\xmin$ shifts towards inner radii. The transition from black to green lines 
corresponds to the appearance of the three critical points; the $\xi,\calR$ values are placed within the triangular region in Fig. 6, where 
there are two minima for $f$. The additional, new minimum of $f$ appears close to the center, but the absolute minimum is still
at larger radii, even if $\xmin$ is decreasing; thus, the green curves show a decreasing trend with $\xi$, and correspond to the positions
of $\xmin$ given by eq. (B13), for $k=0$. As $\xi$ increases further, the external, absolute minimum keeps moving towards the center,
while the value of $f$ corresponding to the inner critical point [$k=1$ in eq. (B13)], keeps decreasing. When the $\xi,\calR$ values hit the
dotted line in Fig. 6, the values of $f$ at the two minima become equal, and then the accretion flow develops  two sonic points.
One point is very close to the MBH, the other is at much larger radii. As $\xi$ increases further, the absolute minimum becomes the 
point nearer to the MBH, and $\xmin$ jumps from the green to the red curve. Further increasing $\xi$, the solution moves out of the triangular 
region in Fig. 6, the absolute minimum of $f$ remains the inner one, until the minimum becomes again only one (and the $\xmin$ is given
again by the black curves). The behavior just described for $\xmin$  also explains the reason why $\xmin$ is so small (similar to that of
classical Bondi accretion) in Fig. 1, left panel: the values of $\calR=10^3$ and $\xi=100$ in Fig. 1 correspond to the red solid curve in 
Fig. 3, i.e. to the regime of very low values for $\xmin$.

The colors of the curves in the right panel reflect the fact that $\ln
\lambdat =f(\xmin)-1/2$, and are thus explained by the behavior of
$\xmin$ described above. The curves for $\lambdat$ are continuous, and
do not show jumps, because $f(\xmin)$ decreases continuously for
increasing $\xi$.  For any fixed $\xi$, at fixed $\calR$, $\lambdat$
increases as $\calR$ increases, i.e., as the galaxy mass increases.

As for the Jaffe model, cases with $\chi<1$ would have very small differences from the curves in Fig. 3, and could be 
distinguished only for very large $\xi$ or small $\calR$, i.e., when the gravity of the MBH dominates.

\begin{figure*}
\hskip 0.5truecm
\includegraphics[height=0.45\textwidth,width=0.45\textwidth]{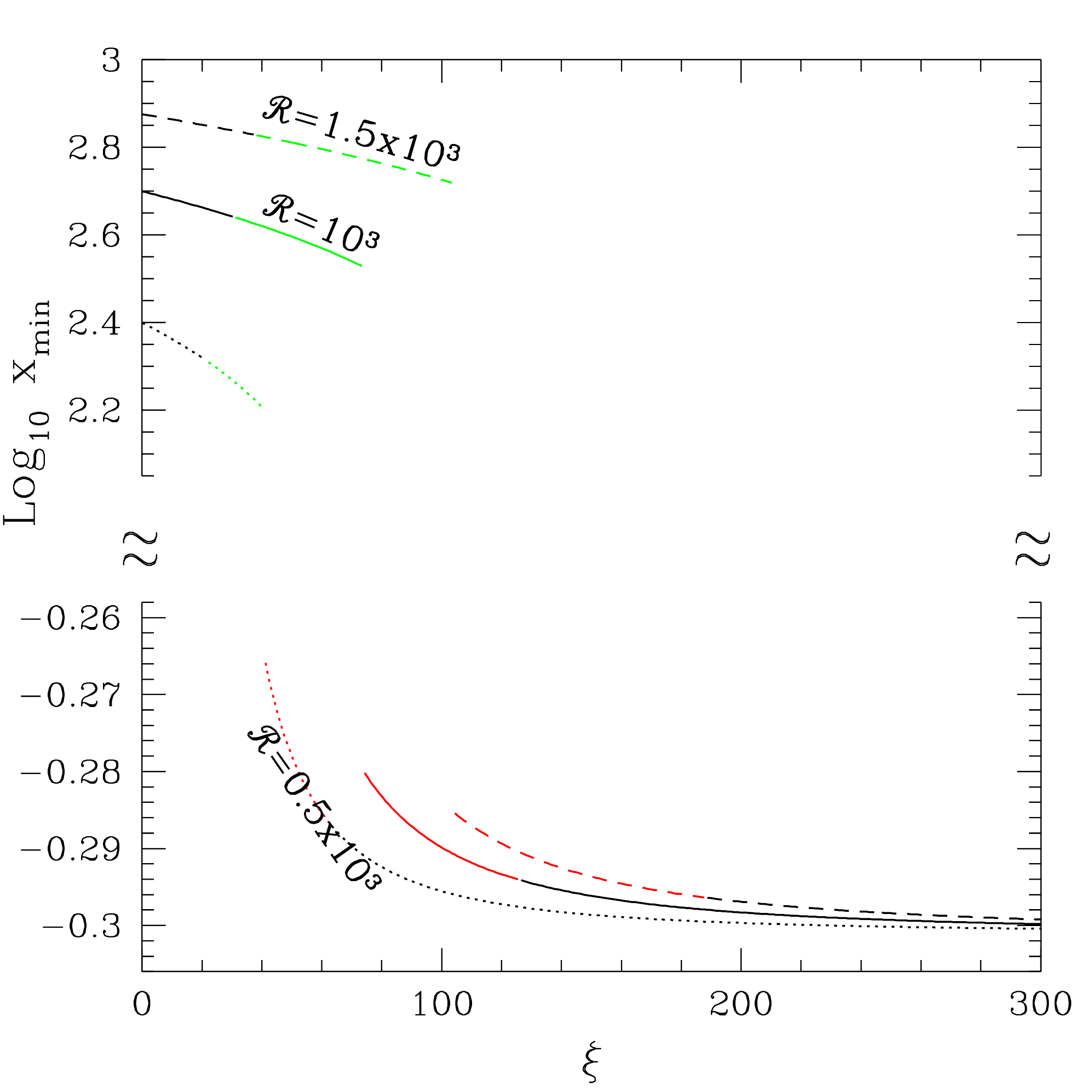}
\hskip 0.8truecm
\includegraphics[height=0.45\textwidth,width=0.45\textwidth]{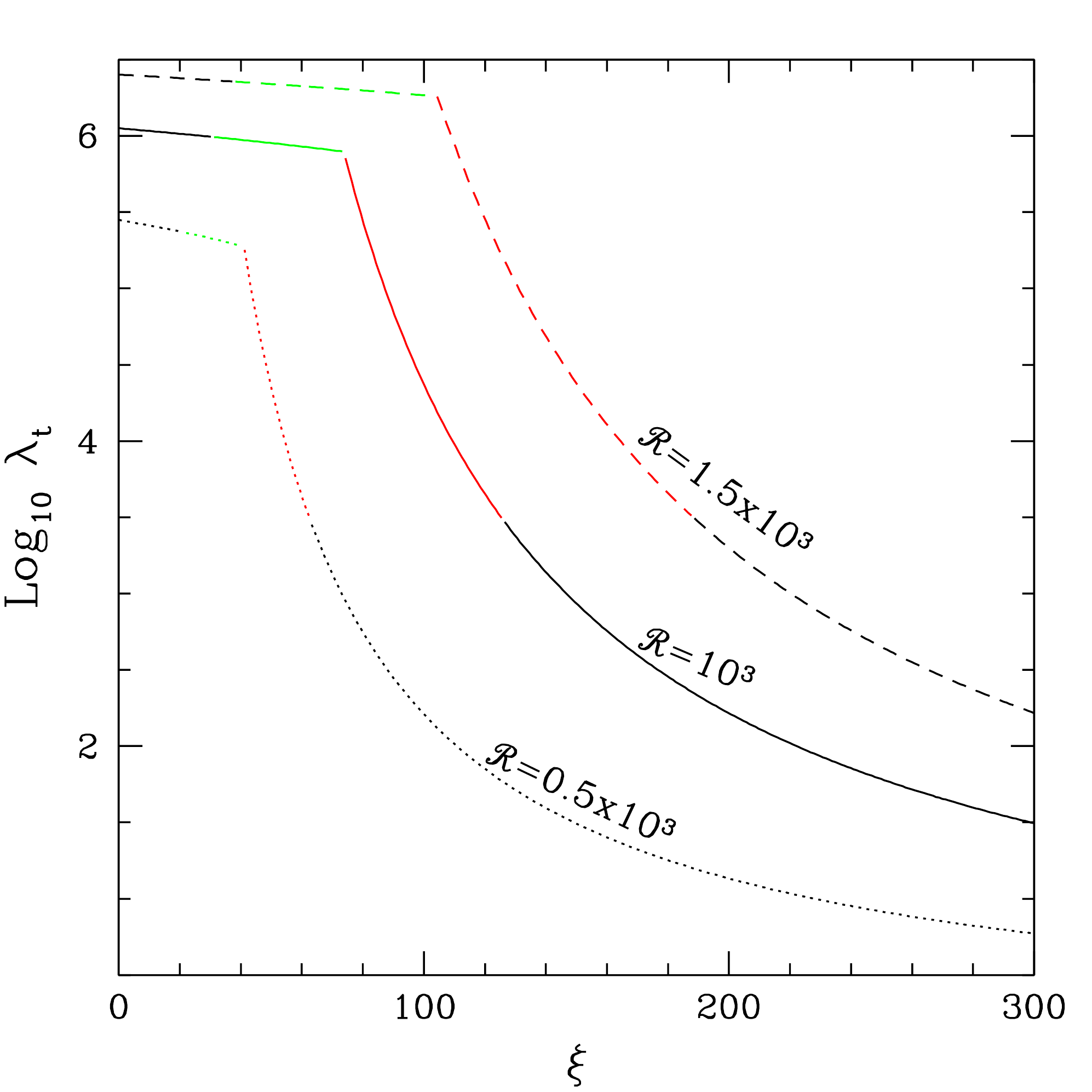}
\caption{Left: position of the sonic point for isothermal accretion in a Hernquist galaxy model, as a function
 of $\xi$, for $\chi=1$. The curves refer to a galaxy mass-to-MBH mass ratio of $\calR=500$
 (dotted), $1000$ (solid), $1500$ (dashed).  The values of $\xmin$
 have been determined using the procedure in Appendix B.
The black lines (at low and large values of $\xi$) correspond to the single minimum (see Fig. 6 in Appendix B, at fixed
 $\calR$); the green and red lines show $\xmin$ when there are two minima, and respectively 
correspond to the solution labeled with $k=1$ and $k=0$ in Appendix
B. The jump from the green to the red solution takes place at  the 
value of $\xi$ such that the two minima of $f$ have the same value
(dotted line in Fig. 6 in Appendix B); for a further 
increase of $\xi$, $\xmin$ switches into the inner region. See Sect. 4.2 for more details. 
Right: the corresponding values of $\lambdat$; note how $\lambdat$ is a continuous function of $\xi$. }
 \label{fig:jaffexmin}
\end{figure*}

\section{An application: evaluation of the mass accretion rate}  
\label{sec:biasclass}

As an application of the previous results, we investigate here the use of the classical
Bondi solution in the interpretation of observational results, in
numerical investigations, or in cosmological simulations involving
galaxies and accretion on their central MBHs (see Sect. 1). In many
such studies, when the instrumental resolution is limited, or the
numerical resolution is inadequate, an estimate of the mass accretion
rate is derived using the classical Bondi solution, taking values of
temperature and density measured at some finite distance from the
MBH. This procedure clearly produces an estimate that can depart from
the true value, even when assuming that accretion fulfills the
hypotheses of the Bondi model (stationariety, spherical symmetry,
etc.).  KCP16 developed the analytical set up of the problem for
generic polytropic accretion, from classical Bondi accretion to the
inclusion of the additional effects of radiation pressure and of an
Hernquist galactic potential; they also investigated numerically the
size of the deviation for some representative cases. Here we
reconsider the problem, for the isothermal case that was
not discussed in detail in KCP16, exploiting our analytical solution,
and extending the investigation also to the case of the Jaffe
potential.

We briefly start with the case of critical accretion in the
classical Bondi problem. For assigned values of $\rhoinf$, $\Tinf$, $\gamma$ and $\Mbh$, the
Bondi radius and accretion rate are given by eqs. (6) and (8):
\begin{equation} 
\rb={G\Mbh\over\csinf^2},\quad	\Mdotb = 4 \pi \rb^2 \lambdacr \rhoinf\csinf.
\end{equation}
If one inserts in eq. (45) the values of $\rho (r) $ and $T(r)$ at
a finite distance $r$ from the MBH, taken along the classical Bondi solution (Sect. 2.1),
and considers them as ``proxies'' for $\rhoinf$ and $\Tinf$, then an {\it estimated}
value of the accretion radius ($\rbe$) and mass
accretion rate ($\Mdotbe$) is obtained:
\begin{equation} 
\rbe (r)\equiv {G\Mbh\over\cs^2(r)},\quad
\Mdotbe(r) \equiv 4 \pi \rbe^2(r)  \lambdacr \rho(r)\cs (r).
\end{equation}
For theoretical investigations  and observational works it is of obvious interest
to know how much these $\rbe$ and $\Mdotbe$  depart from the true values $\rb$
and $\Mdotb$, as a function of $r$, under the assumption that the
Bondi solution holds at all radii.

\begin{figure*}
\hskip -0.2truecm
\includegraphics[height=0.35\textwidth, width=0.35\textwidth]{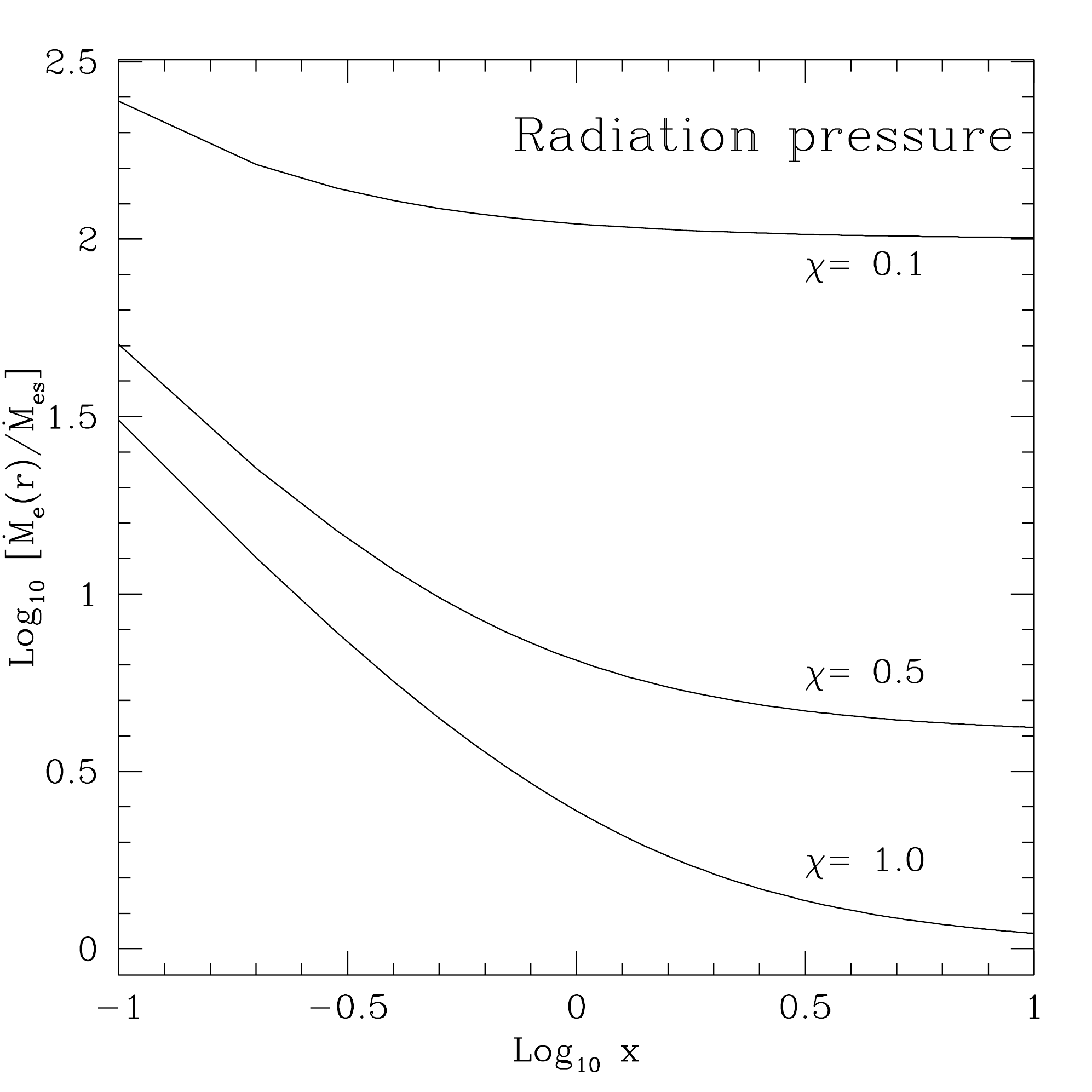}
\hskip -0.3truecm
\includegraphics[height=0.35\textwidth, width=0.35\textwidth]{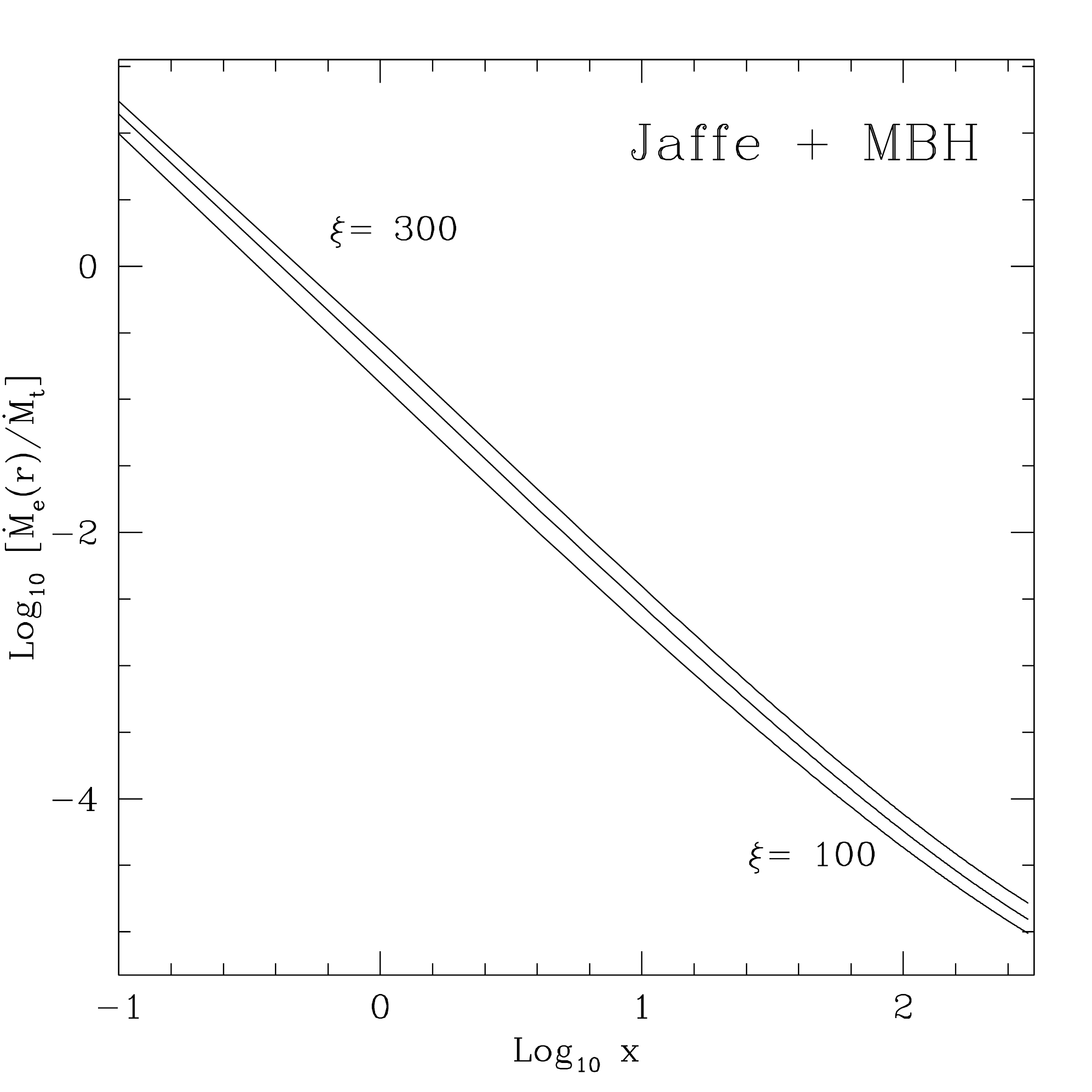}
\hskip -0.3truecm
\includegraphics[height=0.35\textwidth, width=0.35\textwidth]{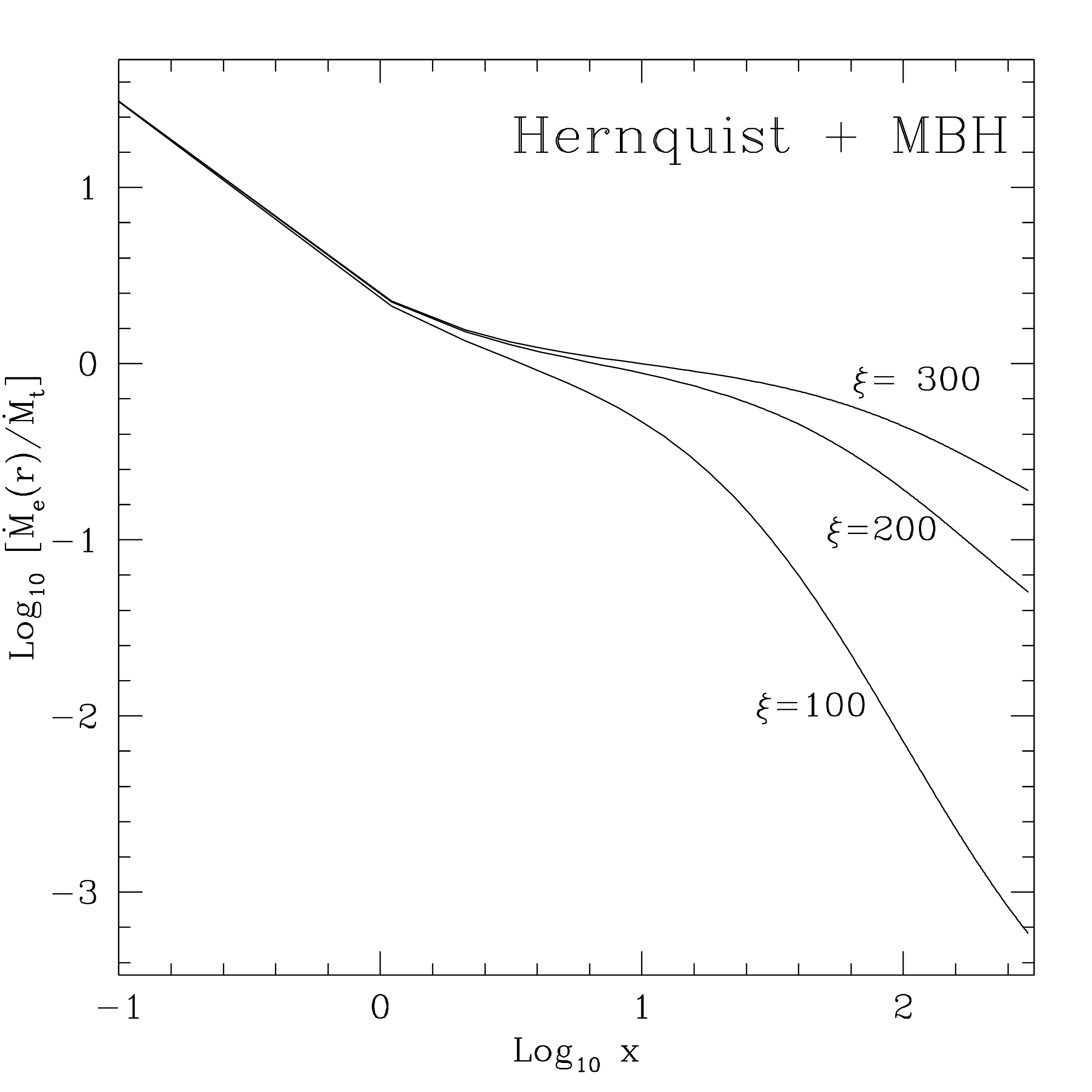}\\
\caption{Ratio between the estimate of the accretion rate $\Mdotbe$ and the true accretion rate 
($\Mdotb$, or $\Mdotbes$, or $\Mdott$), as a function of $x$ (see Sect. 5). Left: classical Bondi 
model with the addition of the effect of electron scattering, for the three indicated values of $\chi$;
$\chi=1$ corresponds to classical Bondi accretion. Middle and right: accretion in  Jaffe and Hernquist
 galaxies, with $\chi=1$, $\calR=10^3$, and for three values of $\xi$. }
\label{}
\end{figure*}

For $\gamma=1$, the isothermal case of present interest, the sound speed is constant, with $\cs (r) =\csinf$, and then $\rbe (r)=\rb$, independently of the
distance from the center at which the temperature is evaluated. Also, $\Mdotbe(r) = 4 \pi \rb^2  \lambdacr \rho(r)\csinf$;
at infinity, the estimate coincides\footnote{Note how, instead, in the monoatomic adiabatic case $(\gamma=5/3)$, one has that ${\Mdotbe}(r)={\Mdotb}$ independently of the distance
from the center, i.e., $\Mdotbe(r)$ does not  deviate from the true accretion rate, while $\rbe(r)$ departs from $\rb$ (KCP16).} with the true value: $\Mdotbe = \Mdotb$. 
At finite radii, from eqs. (45) and (46) the deviation of  $\Mdotbe(r)$ from $\Mdotb$ can be quantified as:
\begin{equation}
{\Mdotbe(r)\over \Mdotb} =\rhotil(x)={\lambdacr\over x^2 \calM(x)},
\end{equation}
where the last identity derives from eq. (20), and $\calM(x)$ is given in eq. (19). 
The deviation then is just given by ${\rhotil(x)} $ at the radius $r$ where the ``measure'' is taken.

Figure 4 shows the trend of $\Mdotbe/\Mdotb$ with $x$ (left panel, for $\chi =1$).
One sees that the use of $\rho(r)$ instead of $\rhoinf$ leads
to an overestimate of the true accretion rate $\Mdotb$ at all $r$; however, 
the overestimate of $\Mdotb$ becomes significant only for $r<\rb$.

We next apply the procedure above to the Bondi problem with radiation
pressure (Sect. 3.1).  Again $\rbe (r)=\rb$, for $\gamma =1$.  We then
quantify the difference between the true [$\Mdotbes = 4 \pi \rb^2
\lambdaes\rhoinf\csinf $, eq. (27)] and estimated [$\Mdotbe (r)$ in
eq. (46)] accretion rate, with $\rho(r)$ now evaluated along the Bondi
solution including the effect of electron scattering.  In this case:
\begin{equation}
{\Mdotbe(r)\over\Mdotbes} = 
{\lambdacr\rhotil (x)\over\lambdaes} = {\lambdacr\over x^2\calM (x)},
\end{equation}
where eq. (29) has been used, and $\calM(x)$ is given by eq. (28).
Thus now the deviation depends not only on the radius $r$ where the density is taken, but also on the value of
$\Mdotbes$ itself, through the parameter $\chi$, that is included in $\calM (x)$ [see eq. (28)].
The trend of $\Mdotbe / \Mdotbes$ as a function of $x$ is shown in Fig. 4, left panel, for two non-special values of $\chi $. 
$\Mdotbe$ overestimates the true accretion rate at all radii,
even by a large factor if $r<\rb$.  Of course, the overestimate increases for decreasing $\chi $, i.e., for increasing radiation 
pressure. The size of the overestimate at large radii is given by eq. (48) and eq. (26), that provide 
$\Mdotbe(r) =\rhotil(x)\Mdotbes/\chi^2$, from which $\Mdotbe/\Mdotbes \to 1/\chi^2$ for $r\to \infty$. 
The behavior of $\Mdotbe/\Mdotbes$ at small radii is discussed below.

Finally, we evaluate the departure of $\Mdotbe(r)$ in eq. (46) from the true mass accretion rate {\it in presence of a galaxy}, given by $\Mdott =4 \pi \rb^2 \lambdat \rhoinf\csinf $ [eq. (34)].  
Taking now $\rho (r) $ along the solution for accretion within the potential of the galaxy and with electron
scattering, one has:
\begin{equation}
{\Mdotbe(r)\over\Mdott}={\lambdacr\rhotil(x)\over\lambdat}={\lambdacr\over
  x^2\calM(x)}.
\end{equation}
Figure 4 shows the radial dependence of $\Mdotbe/\Mdott$ for the Jaffe and Hernquist galaxy models. Both models have the same $\calR=
10^3$ and the same three values of $\xi$; all differences in the trend of their $\Mdotbe / \Mdott$ are then entirely due to the different mass 
density profiles of these two models. The middle and right panels show a clearly different behavior from that of the left panel (no galaxy):
$\Mdotbe(r)$ provides again an overestimate for $r$ taken in the central regions, while $\Mdotbe(r)$ becomes a significant {\it underestimate} 
for large $r$. The position of the radius marking the transition from the region in which there in an overestimate, to that where there is
an underestimate, depends on the specific galaxy model, and on the choice of the galaxy parameter values. An important 
consequence is that, for example, in numerical simulations not resolving $\rb$, $\Mdotbe$ should be boosted by a large factor
to approximate the true accretion rate $\Mdott$. Note how by increasing $\xi$ the effect of the Hernquist galaxy becomes more and more similar to that of a single,
very large mass concentration, and consequently $\Mdotbe$ becomes less and less an underestimate, at large radii. For the Jaffe model this effect is very weak 
for the parameters of Fig. 4.

It is instructive to find the reason for the common trend of $\Mdotbe$ to overestimate the true accretion rate, near the center, 
in all three panels of Fig. 4. This can be achieved from the expansion for $x\to 0^+$ of eq. (33), 
or of the supersonic branch of $W(-1,z)$ in eq. (35). In both ways, for the galaxy models chosen here, and for $\chi >0$,
one has that $\calM (x)\sim {\sqrt{2\chi}}x^{-1/2}$, and so:
\begin{equation}
{\Mdotbe(r)\over\Mdott}\sim {\lambdacr\over\sqrt{2\chi}x^{3/2}}, \quad  x \to 0^+.
\end{equation}
This demonstrates that the effect of the galaxy on the ratio $\Mdotbe/\Mdott$ disappears for $x\to 0^+$; and the trend becomes that of the effect of radiation
pressure, with the respective value of $\chi$. In particular, in Fig. 4 all curves in the central region become similar 
to the curve for $\chi =1$ in the left panel (that corresponds to the effect from the pure MBH). Within the range of $x$ values of Fig. 4
the convergence to the classic ($\chi=1$) Bondi solution has not been reached yet by the Jaffe galaxy,  while instead it has been reached by the 
Hernquist one. This corresponds to the fact that in the central region the galaxy potential is much more important in the Jaffe model than in the 
Hernquist one. Such a difference in the galactic potential is also reflected in the different location of the sonic point $\xmin$ in the two galaxy models: $\xmin$ is very small for the Hernquist
galaxy (see Fig. 3, for the $\xi$ values of Fig. 4), and it is much larger for the Jaffe galaxy (Fig. 2).

For completeness we mention also the case $\chi =0$. While isothermal stationary accretion is impossible in the Hernquist potential (Sect. 4.2),  for the Jaffe model 
one has $\calM (x)\sim 2{\sqrt{(1-\calR/2\xi)\ln(x/\xmin)}}$, for $x \to 0$  (provided that ${\cal R} \geq 2\xi$). From this expression for 
$\calM$ one obtains the analogous of eq. (50).

The expansion of $\calM(x)$ for $x \to \infty$ explains the trend of $\Mdotbe$ at large distances from the MBH. It
gives $\calM (x)\sim \lambdat e^{-(\chi +\calR)/x}/x^2$, and then:
\begin{equation} {\Mdotbe(r)\over\Mdott}\sim {\lambdacr\over\lambdat},
  \quad x \to \infty.
\end{equation}
In the case of a galactic potential, $\lambdat$ becomes very large, and
$\Mdotbe/\Mdott$ correspondingly becomes very small, as shown in
the middle and right panels of Fig. 4. In particular, $\lambdat$ is much larger, and less dependent on 
$\xi$, for the Jaffe than for the Hernquist model (as shown by Figs. 2 and 3). This, together with eq. (51),
explains why in Fig. 4 the underestimate provided by $\Mdotbe$ is much larger for the Jaffe than for the 
Hernquist galaxy;  and why it is hardly dependent on $\xi$, in the Jaffe case, while 
it is largely varying with $\xi$, in the Hernquist one.

\section{Summary and conclusions}

The classical Bondi accretion theory represents the tool commonly
adopted in many investigations requiring an estimate of critical
quantities as the accretion radius and the mass accretion rate; as
examples, we quote here cosmological simulations, when numerical
resolution is not high enough to probe in a self-consistent way the
region near the central MBH, and the interpretation of observational
results, when the instrumental resolution does not allow to reach the
MBH surroundings.  In this work we focus on the case of isothermal
accretion in galaxies with central MBHs, and with radiation pressure
contributed by electron scattering in the optically thin regime.  All
the hypotheses of classical Bondi accretion (stationariety, absence of
rotation, spherical symmetry) were maintained. We show that, notably,
the radial profile of the Mach number can be obtained by using the
Lambert-Euler $W$-function (commonly implemented in computer algebra
systems). Even more remarkably, for the Jaffe and Hernquist galaxies,
also the value of the critical accretion parameter can be analytically
calculated. As an application, we examine the problem of the bias
induced on the estimates of the mass accretion rate by the adoption of
the classical Bondi solution for accretion on a MBH, and of
the gas density and temperature at some finite distance from the
center as proxies for their values at infinity. The main results of this work can
be summarized as follows.

1) For isothermal accretion towards the center of a generic
spherically symmetric potential, and for a given accretion parameter,
the Mach number profile can be written in terms of the $W$-function.
The dependence of the critical accretion parameter from the properties
of the Jaffe and Hernquist galaxy models with a central MBH is given
analytically, even in presence of radiation pressure.

2) For the  Jaffe potential, the determination of the critical accretion parameter involves the solution of a quadratic equation, and there
is only one sonic point for any choice of the parameters describing the galaxy. Moreover, accretion is 
possible even in absence of the central MBH (for a subset of values of the galaxy parameters).

3) For the Hernquist galaxy model, the critical accretion parameter is given by the solution of a cubic equation,
that is fully investigated; one or two sonic points can be present. 
It is also shown that isothermal accretion is not possible in an Hernquist galaxy without a central MBH.
 
4) The structure of the accretion flow is sensitive to the mass density profile of the host galaxy; surprisingly, it turned out to be
quite different even for the two chosen galaxy models, belonging to the class of $\gamma$-models, at fixed total mass and 
scale-length. In fact,  in a Jaffe
galaxy the position of the sonic point depends smoothly on the galaxy properties; on the
contrary,  in the Hernquist potential the flow structure is more complex, and the position of the sonic point can jump from very large to very
small distances from the center, even for a smooth 
variation of the galaxy parameters. For example, for the same (plausible) values of the galaxy parameters, the position of the sonic point is 
$\approx 400\,\rb$ for the Jaffe model, and just $\approx \rb$ for the Hernquist one. 

5) The size of the departure, from the true value, of estimates of the accretion rate $\Mdotbe(r)$ based on
the classical Bondi solution, and the gas properties at a finite distance $r$ from the center, is given as a function of this distance. 
The departure is proportional to the local density of the accreting material.
We derive the formulae for the deviation of $\Mdotbe(r)$ for the three cases of classical Bondi accretion, of accretion with electron scattering, 
and of accretion on a MBH at the center of a galaxy with electron scattering (when the true rate is respectively $\Mdotb$, $\Mdotbes$, and 
$\Mdott$).

6) $\Mdotbe(r)$ is an overestimate of $\Mdotb$, whatever the radius at which the density is taken.
 $\Mdotbe(r)$ is always an even larger overestimate of $\Mdotbes$, when radiation pressure is present; of course, the overestimate becomes 
larger for increasing radiation pressure. In presence of a galaxy, $\Mdotbe(r)$ overestimates $\Mdott$, if the density is taken in the central 
region, and {\it underestimates} $\Mdott$ if it is taken outside a few Bondi radii.
The size of the overestimate, approaching the center, becomes less and less dependent on the galaxy profile, and 
close to that of the classical Bondi problem; the size of the underestimate, instead, is very sensitive to the details of the galaxy profile.
A quantitative explanation for these two trends is given. The position of the transition between the two kinds of departure
depends on the details of the density profile of the galaxy. In conclusion, caution should be exerted when proposing general
recipes for the mass supply rate to the MBH.

Finally, we note that the present investigation can be expanded in an interesting respect, considering the link between the gas 
temperature and the galaxy properties. In fact, in the present study we kept $\csinf$ as a free parameter, while, in a galaxy, the gas 
temperature is likely linked to the depth of the potential well (i.e., it should be close to the virial temperature of the system). This means 
that  $\rb$ should depend on $\rg$ and $\Mg$, which introduces a physical scale in the problem.
Remarkably, the link between the temperature and the potential can be expressed analytically for the two-component (stars + dark matter)  Jaffe models 
with a central MBH (Ciotti \& Ziaee Lorzad, in preparation); this will be the subject of a forthcoming work.




\appendix

\section{A. The Lambert-Euler W function} 
\label{app:A}

The Lambert-Euler function $W(z)$ is a multivalued complex function implicitely defined by 
\begin{equation}
W\exp(W)=z,
\end{equation}
where in general $z$ is a complex variable. As well known, several
trascendental equations can be recast and solved in terms of $W$. For
example, it can be shown that the equation 
\begin{equation}
a\,X^b+c+d\ln X=Y,
\end{equation}
for the non negative unknown $X$, and 
where $a$, $b$, $c$, $d$, and  $Y$ are quantities independent of $X$,  
has the solution 
\begin{equation}
X^b ={d\over ab}W\left[ {ab\over d}\exp{(Y-c)b\over d}\right]. 
\end{equation}
The real determination of $W$ is shown in Fig. 5, and it is made of
two branches, called $W(0,z)$ and $W(-1,z)$. For a general discussion
of the properties of the $W$-function we refer to the classical papers
(e.g., Corless et al. 1996).

The  result above shows that eq. (19) gives the required solution to the problem in Section 2,
as detailed in Sect. 2.1.

\begin{figure}
       \centering 
\includegraphics[height=0.45\textwidth, width=0.48\textwidth]{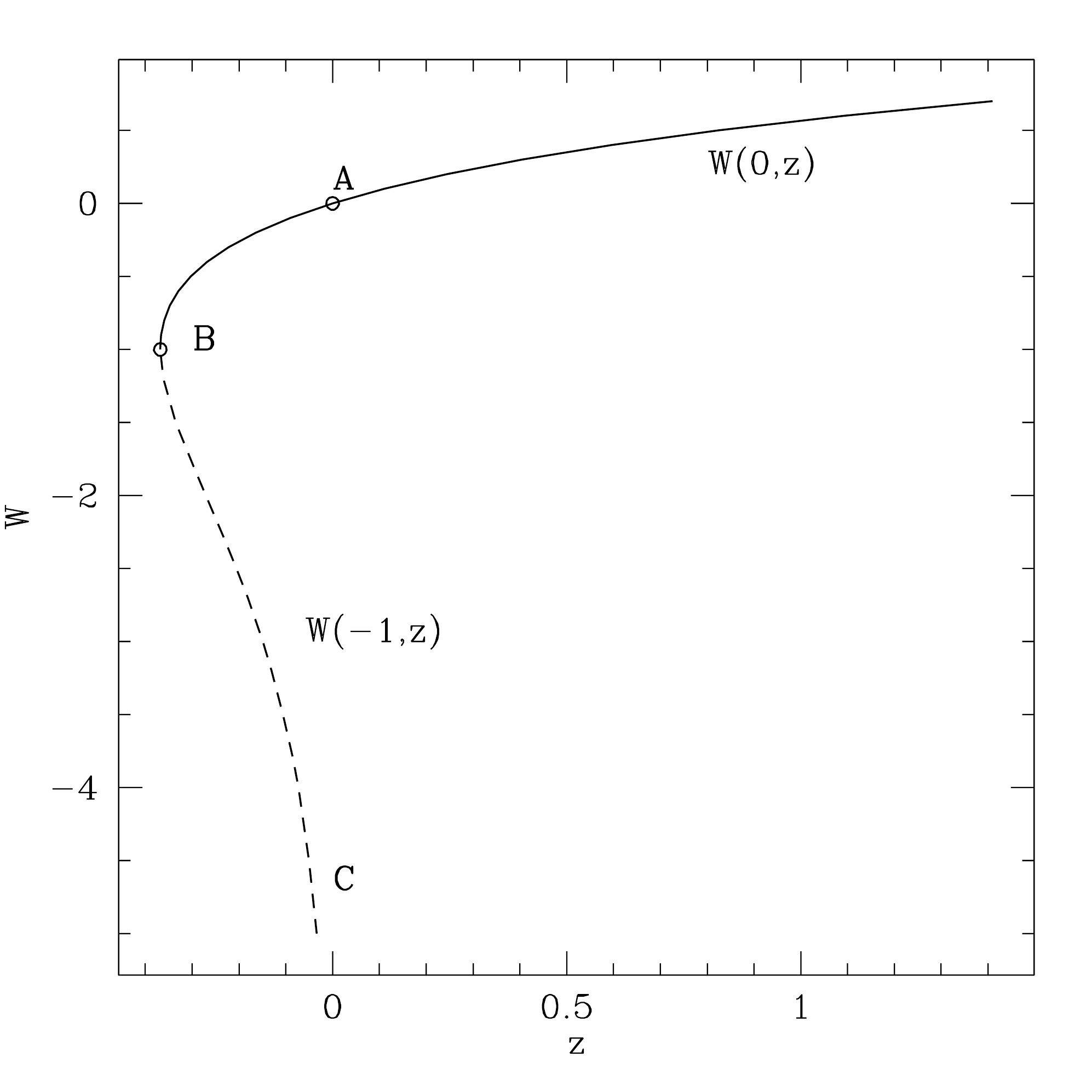}
	\caption{The real determination of the $W$-function. The two branches involved in 
          the solution of the accretion problem are that of 
          $W(0,z)$ (solid line) between the points $A= (0,0)$ and 
          $B=(-1/{\rm e},-1)$, and of $W(-1,z)$ (dashed
          line) between 
          the points $B$ and the asymptotic point $C=(0,-\infty)$.}
\vskip 1truecm
      \label{fig:W}
\end{figure}

\section{B. The critical parameter for isothermal accretion in Jaffe and Hernquist
models with central BH} 
\label{app:Hernquist}

From the discussion in Sect. 2 it follows that the critical value of
the isothermal accretion parameter in the generalized Bondi theory is
given by eq. (15).  Therefore, $\lambdacr$ can be computed explicitly
when $\fmin$ is: in turn, this reduces to the determination of
$\xmin$, i.e. the position of the {\it absolute} minimum of $f$ for $x\geq
0$. In presence of a generic galaxy potential this is not possible
analytically, but in KCP16 it was shown that, quite remarkably, this
can be done for the Hernquist model with a central BH in the general
polytropic case with electron scattering, by solving a cubic
equation. For simplicity the explicit solution was not given, altough
all the needed information to determine the number and position of the
critical points of $f$ were reported and discussed. Here we give the
explicit solution for the Jaffe and Hernquist galaxy models in the
isothermal case and electron scattering.

\subsection{The Jaffe model}

At variance with the Hernquist model, where $\xmin$ (and so $\fmin$
and $\lambdat$) can be computed analytically for $1\leq\gamma\leq
5/3$ (KCP16), in the Jaffe case the explicit expression of $\xmin$ can
be obtained only in the isothermal case, $\gamma =1$. In this case,
however, the situation is much simpler than for the Hernquist
model. In fact, independently of the value of the galaxy parameters,
there is only a single mininum, as can be proved by a
study of eq. (37):
\begin{equation}
f'={2g(x)\over x^2 (\xi+x)},\quad
g=x^2-{\calR +\chi -2\xi\over 2}x -{\chi\xi\over 2}.
\end{equation}

The discriminant of the function $g$ is non-negative for all
physical (i.e., positive) values of
$\calR$, $\xi$ and $\chi$, so that $f'$ has two real
solutions. Moreover, the Descartes' sign rule shows that for $\xi >0$
and $\chi >0$ one solution of $g=0$ is negative and one is positive,
corresponding to the searched position of the minimum of $f$, 
\begin{equation}
\xmin ={\calR + \chi -2\xi + \sqrt{(\calR +\chi -2\xi)^2 + 8\chi\xi}\over 4}.
\end{equation}
In the peculiar case of $\chi =0$ (i.e., when radiation pressure
cancels exactly the gravitational field of the BH, and the effective
potential is only due to the galaxy), a solution of the accretion problem 
is possible only for $\calR\geq 2\xi$, with $\xmin$ and $\lambdat$ given in Sect. 4.1.
When $\calR\leq 2\xi$, the function $g$ in
eq. (B1) is positive for $x >0$, so that the minimum is attained at
the origin.  When $\calR=2\xi$, we have $\xmin =0$ (the sonic point is at the origin),
$\fmin=2\ln\xi$, so that it is {\it finite}, and $\lambdat =\xi^2/\sqrt{e}$, consistently
with the limit of $\lambdat$ for $\calR \to 2\xi$ in eq. (42). For
$\calR<2\xi$, $\fmin= -\infty$, and so no accretion is possibile,
because one would derive $\lambdat =0$. These different behaviors
for the Jaffe model with $\chi =0$ arise because the galaxy potential
is logarithmic, with the possibility to ``compensate'' near the origin
(at least for some choices of $\calR$ and $\xi$) the term $2\ln x$,
even in absence of the gravitational field of the MBH. As we will see, the
more ``regular'' nature of the Hernquist potential at the center makes
this impossible, and no accretion can take place in the isothermal case for
$\chi =0$, independently of the galaxy parameters.

\subsection{The Hernquist model}

In case of isothermal accretion in a Hernquist galaxy with central BH
and radiation pressure due to electron scattering, the critical points
of $f$ are placed at the non-negative zeroes of
\begin{equation}
f'={2g(x)\over x^2(\xi+x)^2},\quad
g=x^3 - {\calR +\chi -4\xi\over 2} x^2 + \xi(\xi-\chi)x - {\xi^2\chi\over 2}.
\end{equation}

The limiting cases of $\xi=0$ and $\chi=0$ are discussed in Sect. 4.2;
here we discuss the more realistic case of $\xi >0$ and $\chi >0$.
The constant term for $g$ in eq. (B3) is strictly negative, while the
coefficient of the cubic term is positive; thus, $f$ has always at
least one minimum for $x>0$.  However, $g$ 
 can have three real zeros, for some particular values of
$\calR$, $\xi$ and $\chi$.  It is therefore important to have
quantitative criteria to determine the number and position of the
critical points of $f$, and in particular of the absolute minimum
$\fmin$ that is required.

The existence and the position
of the zeros of $g$ for assigned values of the parameters can be
determined from the theory of cubic equations, as follows.  With the
substitution
\begin{equation}
x=z+ {\calR +\chi -4\xi\over 6},
\end{equation}
the function $g(x)$ in eq. (B3) reduces to the canonical depressed form $g_c(z)=z^3+pz+q$, with
\begin{equation}
p=-{\calR^2 -2\calR (4\xi-\chi) +(2\xi+\chi)^2\over 12},\quad
q=-{\calR^3 -3\calR^2 (4\xi-\chi) +3\calR (10\xi^2
  -2\xi\chi+\chi^2)+(2\xi +\chi)^3\over 108},
\end{equation} 
so that once the zeroes $z_k$ $(k=0,1,2)$ of $g_c(z)$ are known, the corresponding
solutions $x_k$ of $g(x)$ are obtained from eq. (B4).  As well known
the number of real zeros of a cubic equation with real coefficients is determined by the sign of the
function
\begin{equation}
R\equiv {q^2\over 4}+{p^3\over 27}=
{\xi^2\calR\left[\chi \calR^2 - (\xi^2 + 10\chi \xi -2\chi^2)\calR 
    +(2\xi   + \chi)^3\right]\over 432},
\end{equation}
where the last expression is obtained from eq. (B5). This expression can also be obtained from the more general
eq. (C4) in KCP16, evaluated for $\gamma=1$.

All the roots appearing in the following equations must be intended as arithmetic roots.
When $R>0$ there is only one real solution given by
\begin{equation}
z_0=\sqrt[3]{\sqrt{R}-q/2} -{p/3\over\sqrt[3]{\sqrt{R}-q/2}},
\end{equation}
and so $z_0$ determines the location of
the only minimum of $f$, i.e. $\xmin=x_0$. By inspection of eq. (B6), it follows that $R>0$ both for
small and large values of $\calR$ at fixed $\xi$, and for small and
large values of $\xi$ at fixed $\calR$. The exact boundary of the
region in the $(\xi ,\calR)$ plane corresponding to $R>0$ can be
determined rigorously. In fact, for $\xi<4\chi$ the discriminant of
the quadratic function of $\calR$ in eq. (B6) is negative, and so
$R>0$ independently of the value of $\calR$. For $\xi \ge 4\chi$, in
accordance with Descartes' sign rule, there are two positive values of $\calR$:
\begin{equation}
\calR_{\rm min,max} ={\xi^2 + 10\chi \xi -2\chi^2 \pm 
  \sqrt{\xi(\xi-4\chi)^3}\over 2\chi},
\end{equation}
so that $R<0$ for $\calR_{\rm min}<R<\calR_{\rm max}$, and
$R>0$ outside this range. Therefore, all points outside the
open triangular region in Fig. 6 correspond to models with a single minimum for
$f$, given by eqs. (B4) and (B7). Note that for $\xi \to \infty$, $\calR_{\rm min} \sim 8\xi$ and
$\calR_{\rm max} \sim \xi^2/\chi$. 

Along the two critical curves defined by eq. (B8) in Fig. 6, 
$R=0$ and $g_c(z)$ has a single real root, and a double
(and so also real) root, given respectively by:
\begin{equation}
z_0={3q\over p},\quad z_{1,2}=-{z_0\over 2}.
\end{equation}
The positions of the associated zeros of $g(x)$ are given by a surprisingly simple expression. In fact, along the
lower solid curve in Fig. 6, when $\calR=\calR_{\rm min}$, one has that:
\begin{equation}
0<x_0={\xi\left[\xi-2\chi -\sqrt{\xi (\xi -4\chi)}\right]\over
  4\chi}\leq  x_{1,2}= {\xi + \sqrt{\xi (\xi -4\chi)}\over 2},
\end{equation}
while for $\calR=\calR_{\rm max}$ (the upper solid curve in Fig. 6)
one has that:
\begin{equation}
0<x_{1,2}={\xi - \sqrt{\xi (\xi -4\chi)}\over 2}\leq 
x_0={\xi\left[\xi-2\chi +\sqrt{\xi (\xi -4\chi)}\right]\over 4\chi}. 
\end{equation}
In both cases, being $x_{1,2}$ a double root, $f$ has an {\it
  inflection} point there, and the only minimum is placed at
$\xmin=x_0$ given in eqs. (B10)-(B11).

When $\xi=4\chi$ (and so $\calR=27\chi$) all solutions 
collapse (heavy dot in Fig. 6), and there is only a third-order minimum, placed at 
\begin{equation}
\xmin=x_0=x_1=x_2=2\chi,\quad \fmin=5+2\ln(2\chi) 
\end{equation}
where the last identity follows from eq. (45). Note how
$\fmin\to -\infty$ and $\lambdacr\to 0$ for $\chi\to 0$.

For $R<0$ [when necessarily $p<0$ from eq. (B6)], i.e. for $\calR_{\rm min} < \calR
< \calR_{\rm max}$, there are three real roots of $g_c(z)$, given by:
\begin{equation}
z_k=2\sqrt{-{p\over 3}}\cos\left({\varphi\over 3}+{2\pi k\over
    3}\right),\quad k=0,1,2
\end{equation}
with:
\begin{equation}
0\leq\varphi\equiv\arccos\left(-{q\over 2}\sqrt{-{27\over
      p^3}}\right)\leq\pi .
\end{equation} 
It is simple to prove that $z_1 < z_2 < z_0$; moreover Descartes' sign
rule applied to eq. (B3) with $\calR_{\rm min} < \calR < \calR_{\rm
  max}$ shows that the three real zeros are positive, so that, from eq. (B4), 
$x_1$, $x_2$, and $x_0$ correspond to a minimum, a maximum, and a minimum of
$f$, respectively. In the case $R<0$ we must then determine what is the
position of the absolute minimum among $x_1$ and $x_0$. This is
illustrated by the following procedure. 

\begin{figure}
       \centering 
\includegraphics[height=0.45\textwidth,
width=0.48\textwidth]{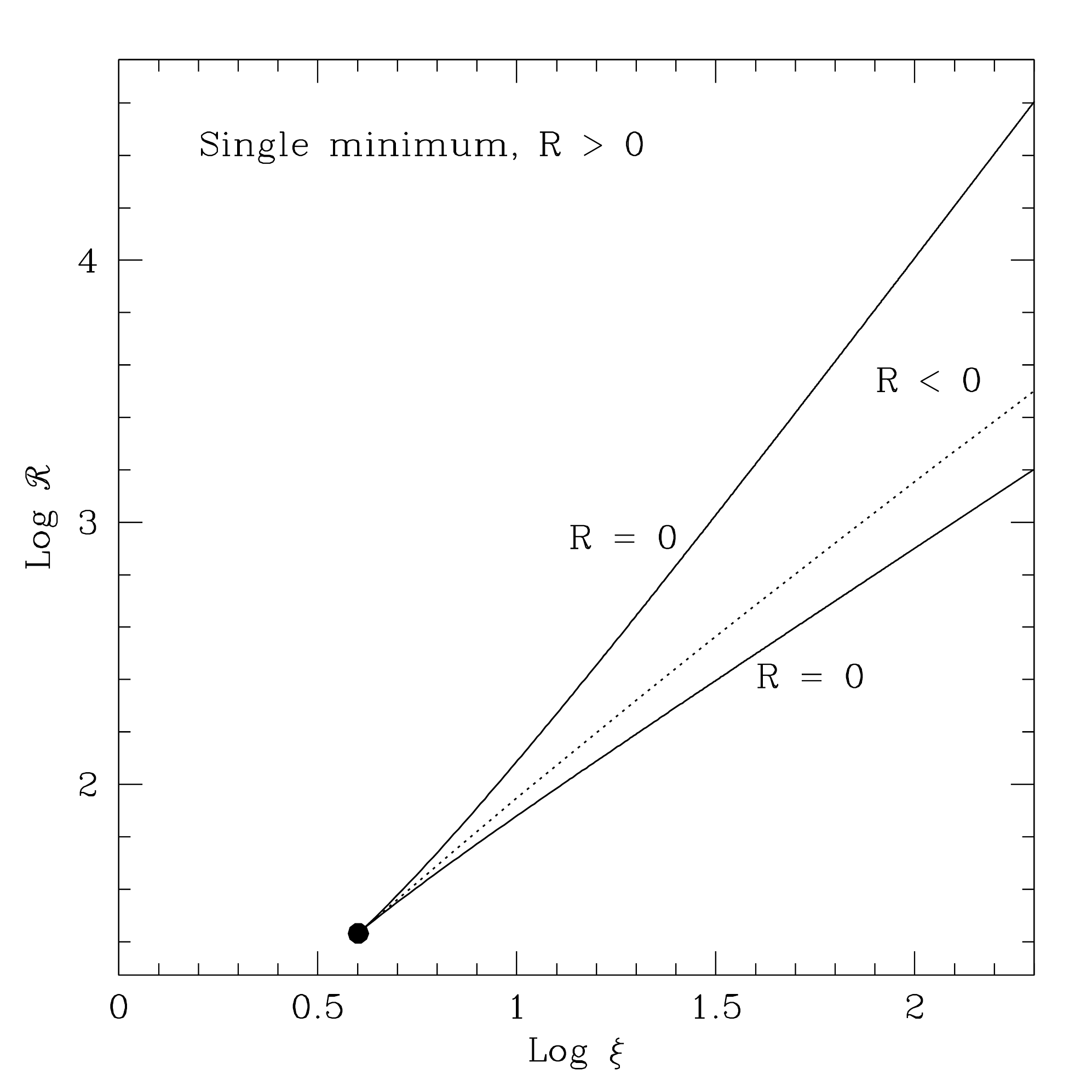}
\caption{Sign of $R$ in eq. (B6) across the $(\calR,\xi)$ plane,  for Hernquist galaxies, in the illustrative
  case $\chi=1$. The two solid lines correspond to eq. (B8), and start at $(\xi,\calR)=(4\chi,27\chi)$.
For $\xi$ and $\calR$ within the triangular region defined by the two solid lines 
 there are two minima and one maximum for $f$ in eq. (45). Along the dotted line there are two sonic points for the 
accretion flow, since the two minima of $f$ have the same depth.}
      \label{fig:HR}
\end{figure}

Suppose we fix a value of $\xi > 4\chi$ and we start increasing
$\calR$ from a very small value, i.e., we move vertically in the
plane of Fig. 6. Initially  there is only one
minimum $\xmin=x_0$ given by eq. (B7), so that $f$ increases
monotonically for $x >x_0$. When $\calR$ reaches the lower curve $R=0$ ($\calR =\calR_{\rm min}$), 
a double root $x_{1,2}$
appears,  corresponding to an inflexion point of $f$, while $\xmin=x_0$,
as in eq. (B10). For $\calR_{\rm min}<\calR<\calR_{\rm max}$, now 
$R<0$, the double root splits, and eq. (B13) applies. The absolute minimum
corresponds to $z_1$, while $z_2$ and $z_0$ correspod to a maximum and a minimum of $f$,
respectively. Increasing further $\calR$, $z_2$ moves towards $z_1$,
while the minimum corresponding to $z_0$ deepens; when the
dotted curve is reached, the two
minima of $f$ have the same value, i.e. $f(x_1)=f(x_0)$, and the accretion flow has two sonic points.  Above the
dotted curve the absolute minimum jumps at $z_0$, and the minimum
at $z_1$ becomes less and less pronounced. When $\calR=\calR_{\rm max}$
the two zeroes $z_1$ and $z_2$ merge, so that $x_{1,2}$ becomes the new
inflection point of $f$, while the absolute minimum is now at
$x_0$, again given by eq. (B7). For higher values of $\calR$, there is
only one minimum given by eq. (B7).

\end{document}